\documentclass[9pt]{article}  \usepackage{times}
\usepackage{graphicx, textgreek}

\usepackage{color}

\usepackage[round,numbers,sort&compress]{natbib} 

\usepackage{amsmath}
\usepackage{nccmath}   
\usepackage{amssymb}
\usepackage{stackrel}
\usepackage{chemarrow}
\usepackage{graphicx}
\usepackage{textgreek}

\usepackage{booktabs}
\usepackage{dcolumn}
\usepackage{rotating}
\usepackage{multirow}

\renewcommand\appendix{\par
	\setcounter{section}{0}
	\setcounter{subsection}{0}
	\setcounter{figure}{0}
	\setcounter{table}{0}
	\renewcommand\thesection{Appendix \Alph{section}}
	\renewcommand\thefigure{\Alph{section}\arabic{figure}}
	\renewcommand\thetable{\Alph{section}\arabic{table}}
}

\begin{document}

	\twocolumn[{\LARGE \textbf{The relationship between the structural transitions of DMPG membranes and the melting process, and their interaction with water\\*[0.2cm]}}
	{\large Thomas Heimburg$^{\,1,\ast}$, Holger Ebel$^{\,2}$, Peter Grabitz$^{\,2}$, Julia Preu$^{\,1}$, and Yue Wang$^{\,1}$\\*[0.1cm]
		{\small $^{1\,}$Niels Bohr Institute, University of Copenhagen, Jagtvej 128, 2200 Copenhagen N, Denmark}\\*[-0.1cm]
		{\small $^{2\,}$Max-Planck-Institute for Biophysical Chemistry, G\"ottingen, Germany - now: MPI for Multidisciplinary Studies}\\*[-0.1cm]
		
		{\normalsize \textbf{ABSTRACT}\hspace{0.5cm} During the melting transition of dimyristoyl phosphatidylglycerol (DMPG), the order of the lipid chains and the three-dimensional, vesicular structural arrangement change simultaneously. These changes result in peculiar heat capacity profiles extended over a broad temperature range with seven $c_p$-maxima. Here, we present calorimetric, viscosity, and volume expansion coefficient data at various ionic strengths and charges. We propose a simple theory that explains the calorimetric data in terms of the coexistence of two membrane geometries, both of which can melt. During the transition, the equilibrium between these two geometries changes cooperatively. This equilibrium depends on the interactions between the membranes and the solvent, on the membrane's charge and the ionic strength of the buffer. Solvent interactions also contribute to the volume change of the membrane phases. Unlike uncharged membranes, we find that enthalpy changes are no longer proportional to volume changes. Therefore, the pressure dependence of the calorimetric profiles differs from that of uncharged membranes. Our theory explains the pressure dependence of calorimetric profiles qualitatively and quantitatively. Furthermore, we demonstrate that the same theory can be used to describe pretransition and ripple formation in phosphatidylcholines. A key takeaway from this article is that solvent molecules (e.g., H$_2$O) are part of the membrane and, in the case of DMPG, water cannot be considered a separate phase.
			\\*[0.3cm] }}
	\noindent\footnotesize{\textbf{Keywords:} structured water ; ripple phase ; pretransition ; relation between enthalpy and volume changes ;\\*[0.1cm]}
	\noindent\footnotesize {$^{\ast}$corresponding author, theimbu@nbi.ku.dk. }\\
	\vspace{0.3cm}
	]

	\normalsize


\section{Introduction}
\label{introduction}

\begin{figure}[htbp]
\centering
\includegraphics[width=225pt,height=152pt]{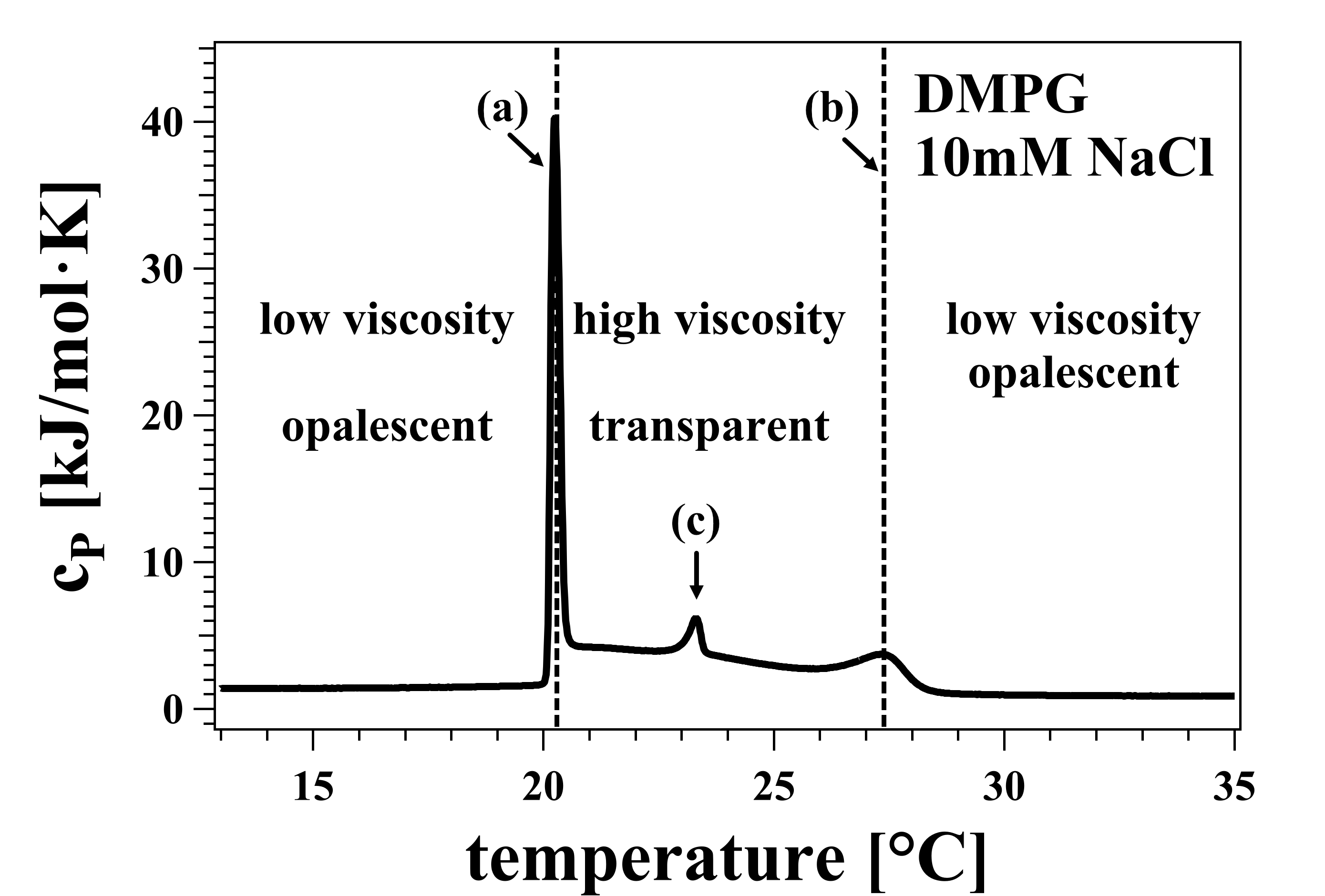}
\caption{The transition profile of DMPG at 10mM NaCl shows a melting regime spanning of about 7--8 degrees with three peaks. The melting is accompanied by changes in viscosity and light scattering. The profile displays sharp peaks at the lower and upper end of the transition (labeled (a) and (b)), and a small peak in the center (labeled (c)).}
\label{fig01_dmpg_intro}
\end{figure}

Lipid membranes can melt from a solid-ordered to a liquid disordered phase. The melting transition represents a change in chain order of the lipids. Such transitions play an important role in bio-membranes \cite{Muzic2019}, for instance, for the emergence of nerve pulses \cite{Heimburg2005c}. The transition is usually continuous, i.e., the susceptibilities do not diverge and both enthalpy, volume and area change continuously upon changes in temperature or pressure. This implies that the extensive variables are monotonously increasing steady functions of the intensive variables. In most reported cases, the transition appears as a single maximum in the susceptibilities, most prominently the heat capacity \cite{Heimburg2007a}, but also the volume and area compressibility, bending elasticity \cite{Heimburg1998} and the capacitive susceptibility \cite{Heimburg2012}. However, there exist interesting deviations from this behavior that did not receive the attention that they deserve. Some transitions display several transition peaks when they are in one way or the other coupled to changes in vesicular morphology. A well-known example are the pre- and main-transition of multilamellar uncharged vesicles that can be seen as two peaks of a continuous melting process coupled to the appearance of periodic ripples on the membrane surface in the melting regime \cite{Heimburg2000}. The pretransition and the main transition represent the lower and upper end of this transition. In the case of dipalmitoyl phosphatidylcholine (DPPC), the range of melting events during which the ripple phase exists is about 7 degrees. A second example is the peculiar melting behavior of dimyristoyl phosphatidylglycerol (DMPG) at low ionic strength. DMPG is a charged lipid. The melting profile shows two or three peaks during a continuous melting process that is also about 7 degrees wide. In the regime between the outer peaks, the lipid dispersion is very viscous and transparent, while it displays a low viscosity and opalescent appearance outside the melting regime. A heat capacity profile of such a transition is shown in Fig. \ref{fig01_dmpg_intro}. Changes in viscosity and opalescence (measured by light scattering) coincide with the changes in the heat capacity profile (see Fig. \ref{fig_1b}) \cite{Heimburg1994}. Such transitions are the topic of this paper.

The integral of the DMPG melting profile corresponds to the total transition enthalpy of the transition, which is similar to that of the lipid dimyristoyl phosphatidylcholine (DMPC), a lipid that displays similar thermodynamic parameters as DMPG. However, DMPC is uncharged (zwitterionic). The coupling between pre- and main transition is usually not taken into account and it is therefore assumed that it displays only one single melting peak. The complex melting behavior of DMPG allows for deep insights into the nature of melting processes.

The unusual behavior of DMPG was first reported by Gershfeld and Nossal \cite{Gershfeld1986}.  Ever since, there have been many publications that describe this behavior \cite{Gershfeld1989, Heimburg1994, Riske1997, Schneider1999, Riske1999, LamyFreund2003, Riske2003, Riske2003b, Kinoshita2008, Fernandez2008, Shen2008, Riske2009, Riske2009b, Spinozzi2010, Alakoskela2010, Barroso2010, Loew2011, Henriques2011, Ito2015, Kelley2020, Schonfeldova2021, Gershfeld2023}. In the past, we proposed that the melting behavior results from changes of the vesicular structure of DMPG that is coupled to changes in the membrane elasticity in the transition regime \cite{Schneider1999}. Such a transition scheme effectively describes two kind of transitions that are not independent of each other: 1. a transition of chain order as described above and 2. a transition in geometry and shape of the macroscopic lipid ensemble. Much of this article is dedicated to understand this coupling on a thermodynamics level.

Even without any theory, there are numerous experimental observations that allow to make informed guesses about what is behind the peculiar behavior of DMPG:

\begin{itemize}
\item The shape of the heat capacity profile depends on ionic strength. At 10mM NaCl, the transition range is about 7 degrees and the relative viscosity (i.e., the normalized viscosity normalized by the viscosity of water at the same temperature) is up to 80. At an ionic strength of 200mM NaCl, the transition profile only displays one single peak, and viscosity changes in the transition are only about 0.05, i.e. hardly visible (see Fig. \ref{fig_1b}). From this can be concluded that the melting behavior is strongly dependent on the electrostatics of the system.

\item Upon increasing the lipid concentration, the melting regime becomes narrower until DMPG behaves like an uncharged lipid such as DMPC \cite{Heimburg1994, Kodama1998}. This has been confirmed for a concentration range of at least 500 \textmu M \cite{Riske2002} mol to 150 mM \cite{Heimburg1994}. In this range, the outer peaks of the transition are further separated the more dilute the sample is. This suggests that the intermediate phase, which is viscous and transparent, interacts with more water than the low and the high temperature phases outside of the melting regime. If the lipid concentration is high, less water is available per lipid. At 150 mM DMPG, there is about 332 water molecules per lipid. At 0.5 mM, there are 111000 water molecules per lipid. Let us assume that DMPG and water possess a density of about 1 g\slash{}ml, that the membrane thickness is 4 nm in the fluid state and that the membrane is flat, this implies that the thickness of a water layer around a membrane is approximately 17 nm on each side of the membrane at 150mM. At 0.5 mM lipid the thickness of the aqueous layer grows to 6 \textmu m. The fact that the calorimetric profiles change in this whole concentration regime from very low to very high concentrations implies that there is no free water and all water molecules interact with the lipid. Loew et al. \cite{Loew2011} called this ``infinite swelling''. Thus, one has to consider the lipid dispersion as one single phase that consists of membranes with very thick interfacial water layers, i.e., a hydrogel where all water is associated to the lipid system. Because, if water was independent of the membrane, and the hydrated membrane phase were separate from an aqueous phase, additional water would not change the heat capacity profiles.

\end{itemize}


\subsection{Proportionality between heat capacity and elastic constants}
\label{proportionalitybetweenheatcapacityandelasticconstants}

We have shown in the past that for many zwitterionic and even biological membranes close to the melting transition, both enthalpy, volume and area are proportional functions of the temperature, with H(T)$\propto$ V(T) $\propto$ A(T) (e.g., \cite{Heimburg1998, Ebel2001, Peters2017}). This is a remarkable experimental fact that allows to relate the susceptibilities with each other:
\begin{eqnarray}\label{eq:Intro01}
\left(\frac{dV}{dT}\right)_p&=&\gamma_V\cdot c_p \quad\mbox{volume expansion coefficient}\nonumber\\
\kappa_T^V(T)&=&\frac{T\cdot \gamma_V^2}{V}c_p \quad\mbox{isothermal volume compress.}\\
\kappa_T^A(T)&=&\frac{T\cdot \gamma_A^2}{V}c_p \quad\mbox{isothermal area compress.}\;, \nonumber
\end{eqnarray}
where $\gamma_V=7.8\cdot 10^{-10}$ m$^2$\slash{}N and $\gamma_A=0.893$ m\slash{}N are parameters for DPPC that assume very similar values for other lipids like DMPC, lipid mixtures and biomembranes from lung surfactant and $E.coli$ \cite{Muzic2019}. We have also shown that the proportional relations imply that after a perturbation there is only one major relaxation process and the relaxation constant can be determined \cite{Grabitz2002}:
\begin{equation}\label{eq:Intro02}
\tau=\frac{T^2}{L}c_p \quad\mbox{relaxation time}\;,
\end{equation}
where $L$ is a phenomenological coefficient, which is $6-8\cdot 10^8$ J$\cdot$K\slash{}mol$\cdot$s for multilamellar vesicles and is $14-20\cdot 10^8$ J$\cdot$K\slash{}mol$\cdot$ for unilamellar vesicles of DPPC and DMPC as measured by pressure perturbation calorimetry. These correlations are of enormous practical value because they allow to determine the elastic constants and the experimental time scales from the heat capacity which is easy to measure. In particular, the heat capacity profiles allow to determine elastic and relaxation behavior in biological membranes. Such calculations require that the relations in eq. (\ref{eq:Intro01}) are correct. However, there exists one class of charged lipids (including DMPG) with peculiar melting behavior, for which it is not clear whether the above relations are in fact true. Having a counter example to eq. (\ref{eq:Intro01}) may challenge the generality of this relation.

The proportional relation between volume and enthalpy can be confirmed by pressure calorimetry. High pressure leads to a shift of the calorimetric profile in a manner that does not affect the shape of the calorimetric profile. The transition enthalpy increases by \cite{Ebel2001}
\begin{equation}\label{eq:Intro03}
\Delta H_{0, p}=\Delta H_{0, p_0}\cdot (1+\gamma_v\cdot \Delta p)\;,
\end{equation}
where $p_0$ is the atmospheric pressure, $T_{p_0}$ is temperature at atmospheric pressure, and $\Delta p$ is the difference between the experimental pressure and atmospheric pressure. The temperature axis at pressure $p=p_0+\Delta p$ is rescaled by \cite{Ebel2001}
\begin{equation}\label{eq:Intro04}
T_{p}=T_{p_0} \cdot(1+\gamma_v\cdot \Delta p) \;.
\end{equation}
When performing the above transformations, one can calculate the heat capacity profiles at pressure $p=p_0+\Delta p$ from the heat capacity profile at $p=p_0$. Vice versa, if the validity of the transformations in eq.(\ref{eq:Intro01}) can be demonstrated experimentally it proves that volume and enthalpy are proportional functions of temperature. This was shown in detail in \cite{Ebel2001}.

We show here that these relations don't hold for DMPG that involve transitions in membrane geometry.


\section{Materials and Methods}
\label{materialsandmethods}

\textbf{Chemicals:} Dimyristoyl phosphatidylcholine (DMPC) and dimyristoyl phosphatidylglycerol (DMPG) in its sodium form were purchased from Avanti Polar Lipids (Birmingham, AL) and used without further purification. Lipids were dispersed in a 2 mM or 5mM Hepes (Sigma), 1 mM EDTA (Fluka) buffer at pH 7.5. The dispersions were incubated at 45$^\circ$--60$^\circ$C for several minutes and vortexed after incubation. Extended lipid phases formed spontaneously. Some experiments were performed in distilled water to keep the ionic strength low. The pH was checked in those experiments.

\textbf{Calorimetry:} We used a VP-differential calorimter (DSC) from MicroCal (Northampton, MA). The DSC scan rate was 5 °C\slash{}hour. For DSC-scans at different pressures, we used a steel capillary inserted into the calorimetric cell. In such experiments, the total amount of lipid in the cell is quite imprecise and we display these data in arbitrary units (a.u.).

\textbf{Viscosity Measurements} Viscosities were measured with a Contraves Low Shear 30 viscometer (Stuttgart, Germany) by using a concentric cylinder pendulum (set 1 cup bob combination).
The viscosity of the lipid dispersion is given relative to the viscosity of water at the same temperature.

\textbf{Volume expansion coefficient:}
The volume expansion coefficient was measured with two different methods:

\emph{Differential densitometry:} Differential densitometry was performed on two coupled DMA 602M cells (Anton Paar, Graz, Austria). One cell was filled with the sample and the other one with the reference buffer. Temperature control by a water bath had an accuracy within 0.01--0.02 degrees. Combining two cells increases the accuracy 20-fold. Details are describe in \cite{Ebel2001}.

\emph{Volume perturbation calorimetry (PPC):} We used the pressure perturbation unit of the VP-calorimeter. The PPC experiment, its design and the theory behind it was described in detail in \cite{Grabitz2002, Heerklotz2002b}. PPC relies on the following principles: At a given temperature, a small pressure perturbation is applied which leads to a release of heat. The heat change is recorded and translated into an excess volume change using the Clausius-Clapeyron equation. This was derived in \cite{Heerklotz2002b}. The experiment is then repeated at all temperatures. Both, DSC and PPC probe the lipid-water system as a whole.

\textbf{Theory:} All of our simulations are based on a simple model assuming an equilibrium between two membrane geometries which are influenced by the melting process (described in \cite{Schneider1999}). Here, this model is refined in order to reflect the asymmetry of the calorimetric profiles and their pressure dependence.


\section{Experimental Results}
\label{experimentalresults}


\subsection{Heat capacity and viscosity as a function of ionic strength and pH}
\label{heatcapacityandviscosityasafunctionofionicstrengthandph}

\begin{figure}[htbp]
\centering
\includegraphics[width=225pt,height=151pt]{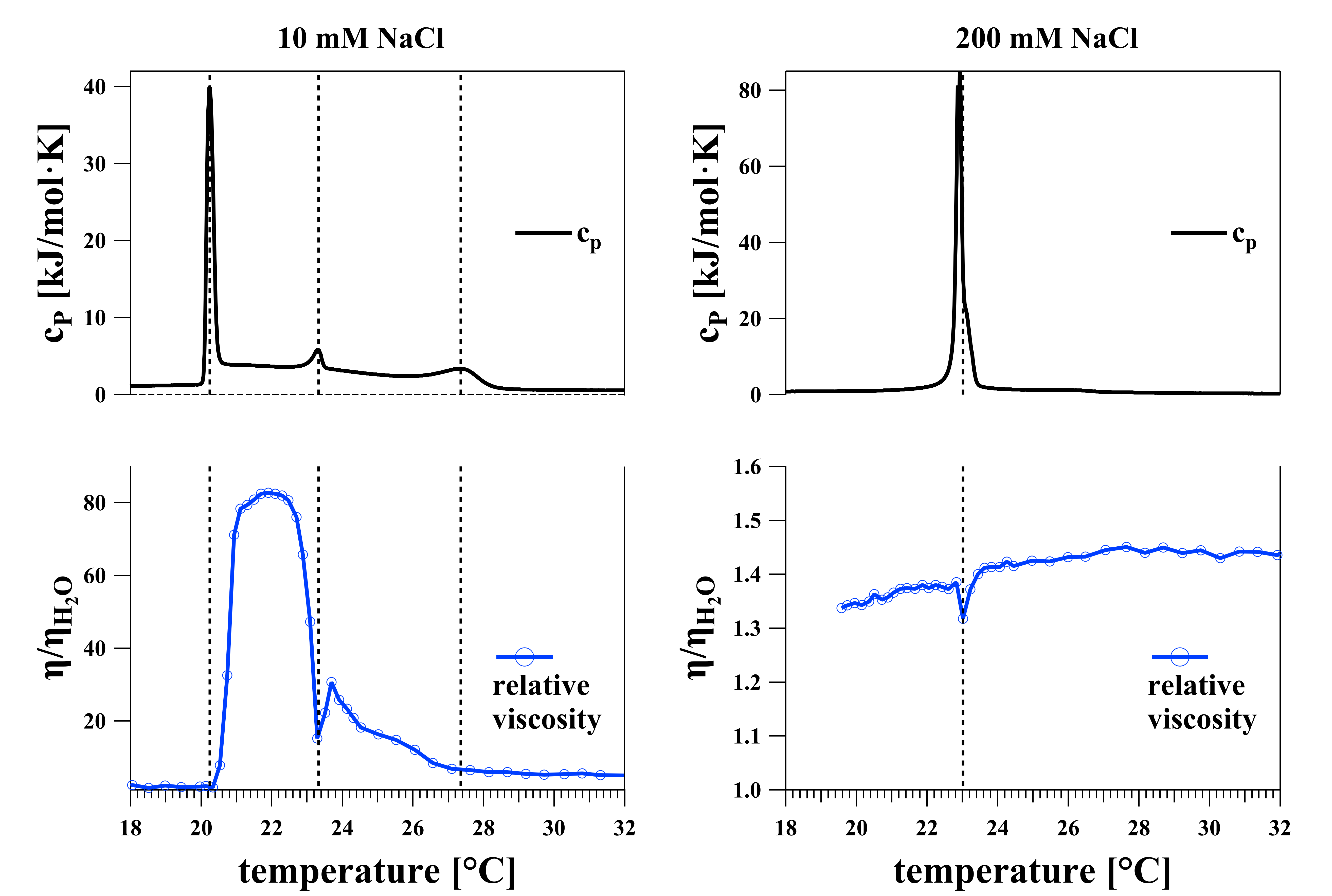}
\caption{Melting of 5mM DMPG at pH 7.8. Left: heat capacity profile and viscosity at 10mM NaCl. Right: heat capacity profile and viscosity at 200mM. The viscosity of the lipid dispersion is normalized by the viscosity of pure water at the same temperature. Data adapted from \cite{Ebel1999}.}
\label{fig_1b}
\end{figure}

The melting of DMPG is highly dependent on both, concentration and charge. The influence of concentration was discussed in the introduction. However, the unusual melting behavior of DMPG also strongly depends on membrane charge, which is influenced by monovalent ions. A high ion concentration leads to screening of the surface charges and to a reduction of the electrostatic surface potential \cite{Trauble1976}. Fig. \ref{fig_1b} shows the difference in the calorimetric profile of a 5mM DMPG dispersion at 10 mM NaCl (Fig. \ref{fig_1b}, left) and at 200 mM NaCl Fig. (\ref{fig_1b}, right). One finds a transition width of about 8 degrees from $\sim$20$^\circ$ to $\sim$ 28$^\circ$C for the 10mM NaCl sample, while it is less than 1 degree at 23$^\circ$C at 200mM NaCl - very similar to uncharged vesicles of DMPC. The viscosity of the lipid dispersion is more than 80 times larger than that of water at 22$^\circ$C at low salt, while one finds only a very minor decrease in viscosity at 23$^\circ$C in the high salt sample. This makes it clear that the melting events in DMPG dramatically depend on surface potential.

\begin{figure}[htbp]
\centering
\includegraphics[width=225pt,height=118pt]{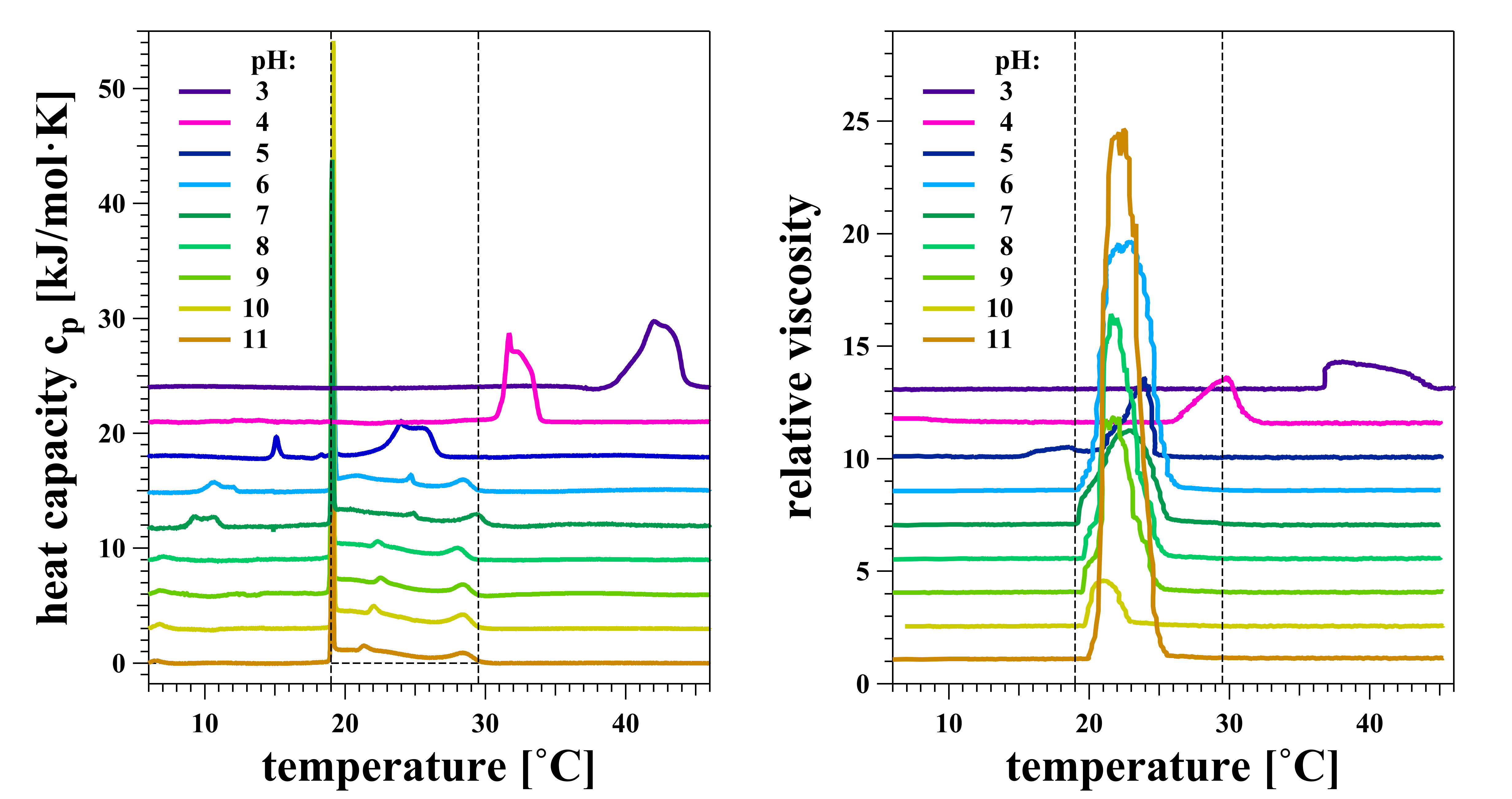}
\caption{Comparison of heat capacity profiles and relative viscosity of 5mM DMPG in 10 mM Hepes , 10 mM NaCl at pH values between 3 and 11. Left: Heat capacity. Right: Relative viscosity. The behavior in the range between pH 6 and 11 is very similar. Below pH 6, the lipids start getting protonated (see \cite{Trauble1976} for explanation). The vertical dashed lines indicate the lower and upper end of the melting regime between pH 6 and 11. Both, heat capacity and viscosity changes are shifted relative to each other for better visibility. Outside the transition regime, $\Delta c_p$ is close to zero and $\eta/\eta_{water}$ is close to 1 for all samples. Data adapted from \cite{Preu2010}.}
\label{figure_julia_viscosity}
\end{figure}

The NaCl concentration given here is that of the added salt. Ions such as sodium and chloride screen fixed charges but do not bind to them. In Fig. \ref{fig_1b} one can see that the melting range is narrower in the presence of NaCl, but the center temperature does not change dramatically. Since the mean melting temperature of DMPG does not change as a function of salt concentration, we assume that sodium is always dissociated. Protons, in contrast, neutralize the changes and abolish charge effects. Here, protons associate with the DMPG head-groups at low pH. This renders the lipids uncharged and the transition temperature increases to much higher values.
Thus, the effect of sodium and of protons on charged membranes is very different although both are single charged cations. This can be seen in the heat capacity profiles of DMPG at different pH values between 3 and 11 as shown in Fig. \ref{figure_julia_viscosity} (left). While the $c_p$-profile is practically unchanged in a pH range between 11 and 6, the center temperature of the transition profiles increasingly moves upwards at pH 5, 4 and 3. At the same time, the width of the transition gets smaller and the relative change in viscosity in the transition regime decreases dramatically. This indicates a protonation of the membrane that is different from the effect of NaCl.

Together with the concentration dependence discussed in the introduction, the results of this section suggest that the strange melting behavior of DMPG is associated to long range ordering of water such that no free water is present. It is aided by surface electrostatic. Shielding or neutralizing of charges abolishes the peculiar melting behavior. So, we find the surprising result that water can be ordered by charged surfaces over a range of several micrometer and larger, i.e., over much larger distances than the thickness of the membrane (up to 3 orders of magnitude larger than the membrane thickness). It should be noted that this result is inferred from infinite lipid phase swelling. Direct experimental evidence, such as relaxation time measurements of water, is still lacking.


\subsection{Violation of the proportionality for DMPG}
\label{violationoftheproportionalityfordmpg}

The proportionality of volume and area in membrane transitions of zwitterionic membranes is an important and valuable finding \cite{Heimburg1998, Ebel2001, Heimburg2007a, Muzic2019}. As discussed in the introduction, it allows to calculate volume and area compressibility, the membrane bending elasticity and the relaxation time scales from the heat capacity. We show below that DMPG does not obey the relations shown in eqs. (\ref{eq:Intro01}), (\ref{eq:Intro03}) and (\ref{eq:Intro04}). When chain melting is coupled to changes in the membrane geometry associated to the melting events. These coupled transitions are characterized by displaying more than one transition peak and a pressure dependence that deviates from that given in eq. (\ref{eq:Intro04}).

\begin{figure}[htbp]
\centering
\includegraphics[width=169pt,height=154pt]{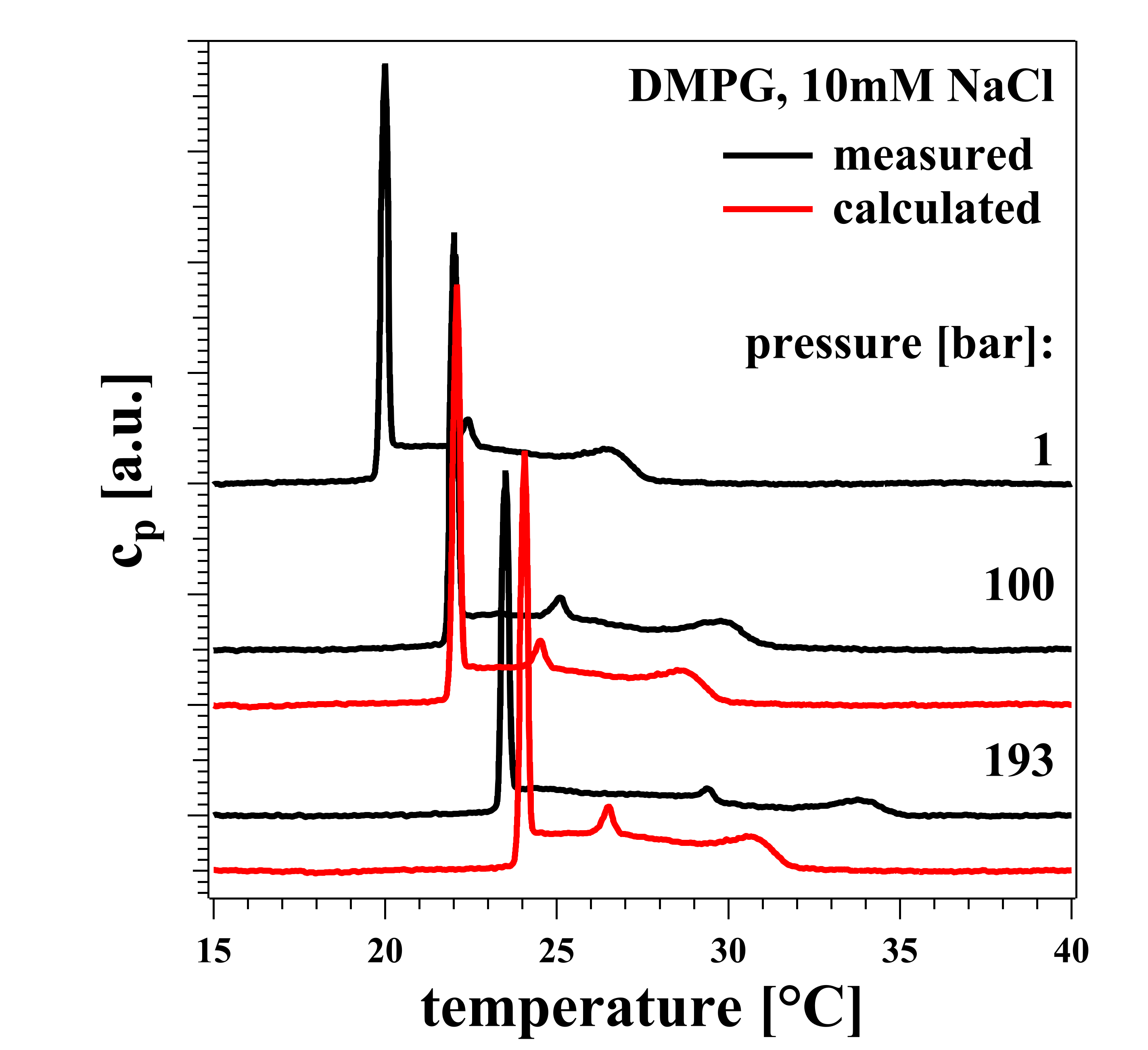}
\caption{Comparison of the pressure dependence of the heat capacity profiles of a DMPG dispersion at 10mM NaCl at three pressures, and the profiles calculated from the heat capacity at 1 bar using relation eq. (\ref{eq:Intro04}). The calculated profiles differ significantly from the measured profiles. Measured data adapted from \cite{Grabitz2001}.}
\label{dmpg_pressure_comp_exp_calc}
\end{figure}

Eqs. (\ref{eq:Intro01}), (\ref{eq:Intro03}) and (\ref{eq:Intro03}) tell us that one can calculate a heat capacity profile at pressure $p$ from the heat capacity profile at atmospheric pressure $p_0$ as long as enthalpy and volume are proportional functions of temperature. Fig. \ref{dmpg_pressure_comp_exp_calc} shows the heat capacity profile of DMPG at 1 bar (top trace), and the heat capacity profiles at measure at 100 bars and 193 bars (black traces) and calculated from this profile (red traces) using eqs. (\ref{eq:Intro03}) and (\ref{eq:Intro04}). We recognize that the red and black profiles are not identical. The measured profiles display a transition width that is significantly larger than the calculated profiles. At 100 bar, the width of the transition is 9.4$^\circ$ experimentally (black trace) and 6.2$^\circ$ for the calculated profile (red trace). At 193 bar, the width of the transition is 11.9$^\circ$ experimentally (black trace) and 8.1$^\circ$ for the calculated profile (red trace). Furthermore, the lower end of the transition is at a lower temperature for the experimental profile as for the calculated profile, and the upper end of the transition is at a higher temperature as for the calculated profile.
This implies:

\begin{enumerate}
\item Enthalpy and volume cannot be proportional functions

\item A proportional relation overestimates the volume change at the lower end of the transition, and underestimates it at the upper end of the transition.

\end{enumerate}

\begin{figure}[htbp]
\centering
\includegraphics[width=225pt,height=121pt]{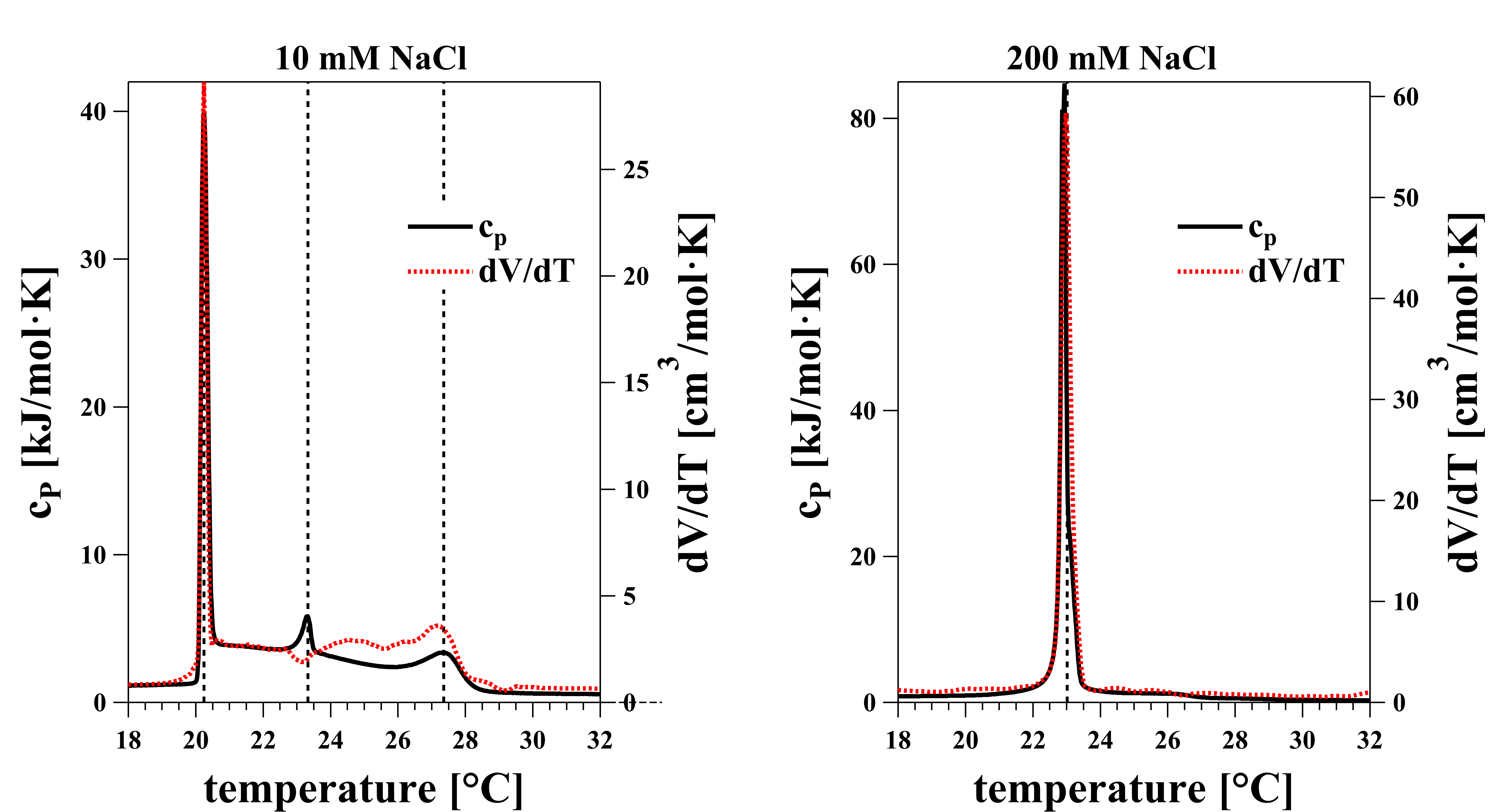}
\caption{Comparison of the heat capacity ($c_p$) and the volume expansion coefficient $\alpha$ measured in a vibrating tube densitometer. 10 mM DMPG were dissolved in a buffer with 5mM HEPES, 1mM EDTA at a pH of 7.8. Left: 10 mM NaCl. The volume expansion coefficient and the heat capacity display a similar but not identical profile. Right: 200 mM NaCl. Data from \cite{Ebel1999}.}
\label{figure_ebel_dmpg_cp_alpha}
\end{figure}

This can be checked in densitometry and pressure perturbation experiments. Fig. \ref{figure_ebel_dmpg_cp_alpha} shows vibrating capillary densitometry experiments on DMPG at 10 mM NaCl (left) and at 200 mM NaCl (right). Besides the fact that the shapes of the profiles are very different, one can recognize that in the left hand panel, the volume expansion coefficient $\alpha\equiv (dV/dT)_p$ and the calorimetric profile do not exactly overlap, which indicates that volume and enthalpy are not proportional functions of temperature. As suggested above from the pressure dependence of the calorimetric profiles, the volume expansion coefficient lies below the $c_p$-profiles at the lower end of the transition range, while it lies above the $c_p$-profile at the upper end of the transition range. The right hand panel shows the same lipid sample at 200 mM NaCl, where any deviations between heat capacity and volume expansion coefficient are less obvious.

\begin{figure}[htbp]
\centering
\includegraphics[width=225pt,height=91pt]{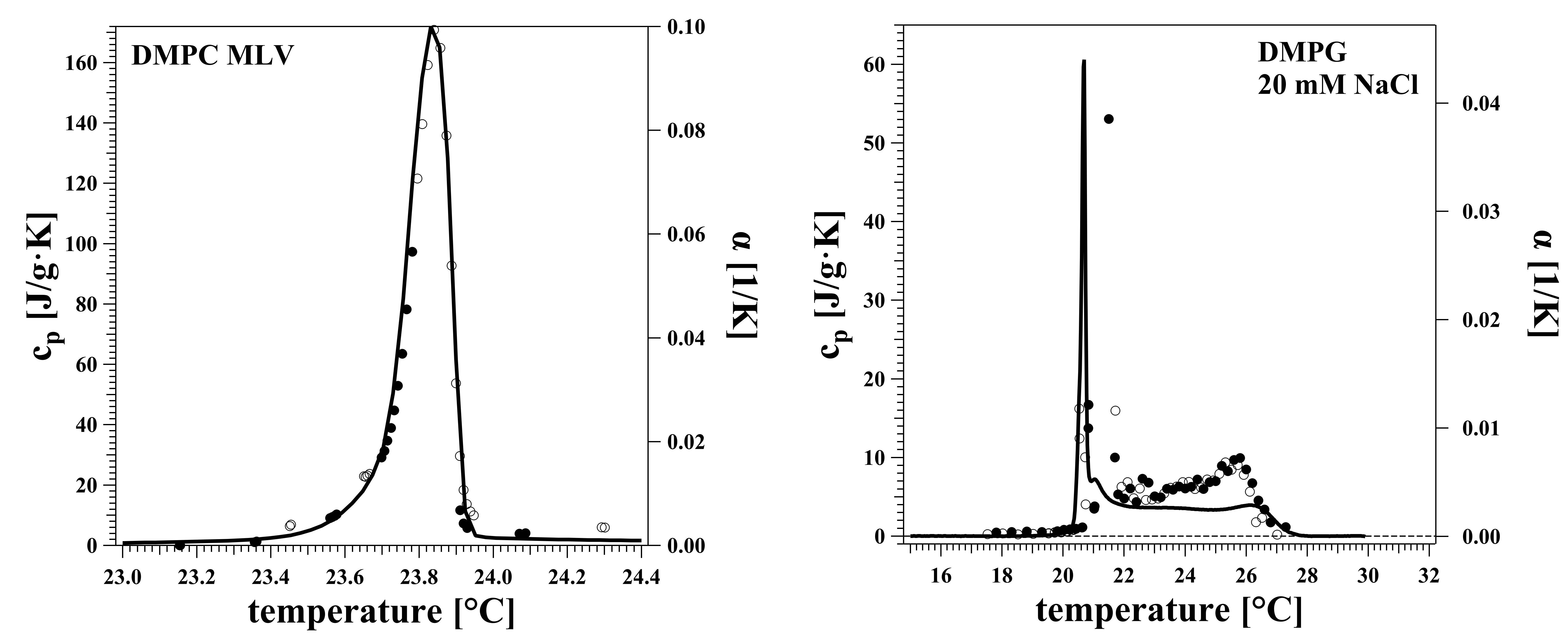}
\caption{Comparison of heat capacity and volume expansion coefficient as measure by pressure perturbation calorimetry. Left: 10 mM DMPC MLV in water. Right: 10 mM DMPG in 20mM NaCl, 5mM Hepes, pH 7.4. While for the uncharged DMPC, heat capacity and volume expansion coefficient display exactly the same temperature dependence, this is not exactly the case for the charged DMPG at low ionic strength. }
\label{figure_ppc_dmpc_dmpg}
\end{figure}

Fig. \ref{figure_ppc_dmpc_dmpg} shows a similar experiment performed with pressure-perturbation calorimetry (PPC) (see \cite{Heerklotz2002b} for description of the experiment). The left had panel shows DMPC multilamellar vesicles (MLV). The volume expansion coefficient and the heat capacity are perfectly superimposable. The right hand panel shows a 10mM DMPG dispersion where it is obvious that heat capacity and volume expansion coefficient deviated in the same manner as in Fig. \ref{figure_ebel_dmpg_cp_alpha}. So, we can state that while heat capacity and volume expansion coefficients are still qualitatively similar, the are quantitatively distinct.

What we see is that in the transitions that involve structural changes caused by interactions of charged membrane surfaces with the solvent (such as in DMPG), the proportionality of heat capacity and volume expansion coefficient is violated, while all samples that do not involve structural changes caused by solvent interactions, the two functions are nearly exactly proportional.


\section{Theory}
\label{theory}


\subsection{Melting equilibria}
\label{meltingequilibria}

In the following, we derive a theory that explains the peculiar shape of the melting profile of DMPG, the deviations of volume expansion from heat capacity and the pressure dependence. Let us assume that two geometries of a lipid membrane exist. They may both melt from a gel state to a fluid state upon changing temperature or pressure. We describe the melting process with a simple thermodynamic model that contains four states as shown in Fig. \textbackslash{}ref\{diagram\_e\}.

\begin{figure}[htbp]
\centering
\includegraphics[width=225pt,height=117pt]{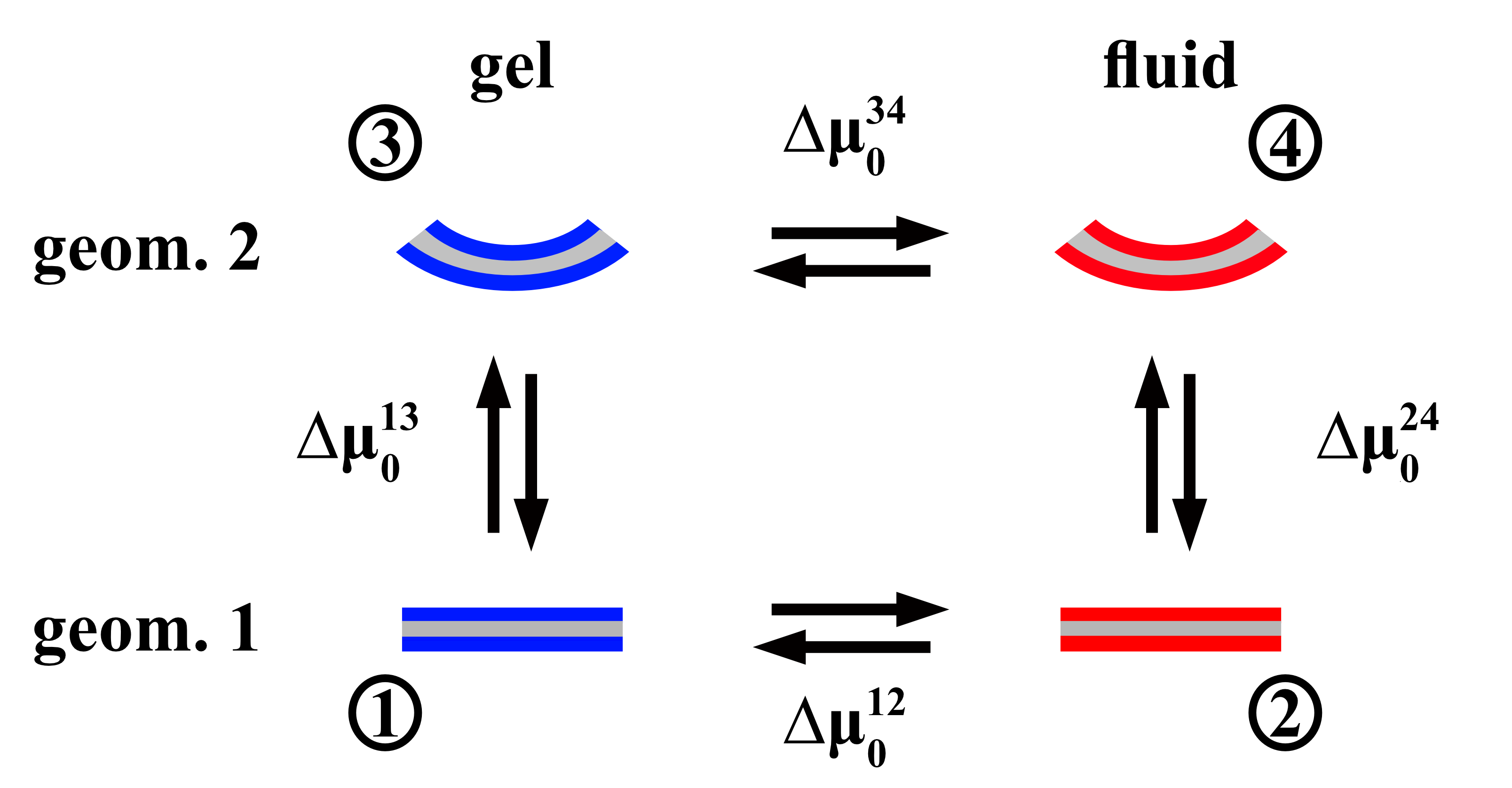}
\caption{In our derivations we assume that there are two geometries, that both can melt from a gel to a fluid state. This yields four states (denoted by 1,2,3 and 4) that differ by temperature and pressure dependent contributions to the chemical potential caused by interactions with the solvent.}
\label{diagram_e}
\end{figure}

The equilibria between the four states display chemical potential differences, $\Delta \mu_0^{12}$, $\Delta \mu_0^{13}$, $\Delta \mu_0^{24}$, and $\Delta \mu_0^{34}$, where
\begin{equation}\label{eq:Theory_A1}
\Delta \mu_0^{ij}(p,T)= \Delta H_0^{ij}+ p\Delta V_0^{ij}-T\Delta S_0^{ij}\;,
\end{equation}
where $\Delta H_0^{ij}$, $\Delta S_0^{ij}$ and $\Delta V_0^{ij}$ denote the enthalpy, entropy and volume differences at atmospheric pressure. They are assumed to be constants. $p$ is the difference of the experimental pressure and atmospheric pressure $p_0$, and $T$ is the absolute temperature.
The equilibrium between states $1$ and $2$ describe the melting of geometry 1, with $T_m^{12}=\Delta H_0^{12}/\Delta S_0^{12}$. Likewise, the equilibrium between states $3$ and $4$ describes the melting of geometry $2$ with $T_0^{34}=\Delta H_0^{34}/\Delta S_0^{34}$. The enthalpy, entropy and volume changes in this transition will be similar to those of the $12$-equilibrium (the equilibrium) with an extra small solvent contribution that changes the precise melting temperature $T_m^{34}$. The dependence of the calorimetric profiles on lipid concentration and protonation of the charged lipids suggests that the different geometries interact differently with the aqueous solvent. Thus, equilibrium $13$ of the gel membrane in the two different geometries in the gel state will be dominated by interactions with the solvent, such that state $3$ interacts better of worse with the solvent than state $1$. Since the free energy is a function of state (the integral of $\mu$ is independent of path), it is clear that $\Delta \mu_0^{24}=\Delta \mu_0^{12}+ \Delta \mu_0^{24}-\Delta \mu_0^{12}$. This implies that the equilibrium $24$ is defined by the other three equilibria and also contains solvent-dependent contributions. The complete diagram is described by three enthalpy-, three entropy- and three volume-differences. Below we will simplify the calculations and assume the some of the entropy and volume differences can be neglected. We will show that with relatively few parameters one can both describe the solvent and the pressure dependence of the calorimetric profiles with surprising qualitative and partially quantitative agreement.

Let us first describe the melting transition of geometry 1:, $\Delta H_0^{12}$ is the melting enthalpy that can be determined in a calorimeter, $\Delta S_0^{12}=\Delta H_0^{12}/T_m^{12}$ (where $T_m^{12 }$ is the melting temperature corresponding to the maximum of the heat capacity), and $\Delta V_0^{12}$ is the volume change in the transition. The latter can be obtained from the pressure dependence of the melting transition as describe above. DMPG at high salt behaves very similar to DMPC lipids \cite{Heimburg1999}. Therefore, we take values similar to those of DMPC: $\Delta H_0^{12}=25$ kJ\slash{}mol, $T_m^{12}=296.15$ K (23$^\circ$C), and $\Delta V_0^{12} = \gamma_V \Delta H_0^{12}$, with $\gamma_V=7.23\cdot 10^{-10}$ m$^2$/N \cite{Ebel2001} (see above).

The transition between gel and fluid is described by an equilibrium constant,
\begin{equation}\label{eq:Theory_A2}
K_{12}(p,T)=\exp\left(-n_{12}\frac{\Delta \mu_0^{12}(p,T)}{RT}\right) \;,
\end{equation}
where $n_{12}$ is a cooperative unit size. It determines the width of the melting transition. (e.g., \cite{Heimburg2007a, Wang2018}). The cooperative unit is the number of lipids that undergoes a transition together. Here, $n_{12}\Delta \mu_0^{12}$ is the free energy per mole of cooperative units, while $\Delta \mu_0^{12}$ is the free energy per mol of lipids. It mimics an effective domain size in the transition. A typical number for unilamellar vesicles of uncharged lipids is $n=120$. The fraction of the membrane in the fluid and in the gel state state is given by
\begin{eqnarray}\label{eq:Theory_A2b}
f_{12}(p,T)&=&\frac{K_{12}(p,T)}{1+K_{12}(p,T)}\quad\mbox{fluid fraction} \;,\nonumber\\
1-f_{12}(p,T)&=&\frac{1}{1+K_{12}(p,T)}\quad\mbox{gel fraction} \;,
\end{eqnarray}
respectively. In the following, we will skip the notation $(p,T)$ for brevity, but for all functions below that include $\Delta \mu_{ij}$, $K_{ij}$ or $f_{ij}$ it is tacitly assumed that they depend on pressure and temperature.

The enthalpy of the transition as a function of temperature is the fluid fraction time the enthalpy change in the transition:
\begin{equation}\label{eq:Theory_A3}
\Delta H_{12}=f_{12}\cdot (\Delta H_0^{12}+p\Delta V_0^{12})\;.
\end{equation}
Equally, the volume as a function for temperature is given by
\begin{equation}\label{eq:Theory_A4}
\Delta V_{12}=f_{12}\cdot \Delta V_0^{12}\;.
\end{equation}
The molar heat capacity is the temperature-derivative of the enthalpy, $\Delta c_p^{12}=(dH_{12}/dT)_p$ (see eq. (\ref{eq:Theory_A3})), and the temperature dependence of the volume can be derived from eq. (\ref{eq:Theory_A4}).:
The entropy as a function of temperature can be obtained by integrating $c_p/T$:
\begin{equation}\label{eq:Theory_A6}
\Delta S_{12}=\int_{T_0}^{T_1} \frac{\Delta c_p^{12}}{T} dT\;,
\end{equation}
where $T_0$ is a temperature below the transition, and $T_1$ is a temperature above the transition. Now, $\Delta \mu_0^{12}(p,T)$ can be calculated by using eq. (\ref{eq:Theory_A1}).

We show below that this description leads to quite satisfactory profiles of the heat capacity very similar to experimental profiles of DMPC. The description contains four parameters, $\Delta H_0^{12}$, $\Delta S_0^{12}$, $\Delta V_0^{12}$ and $n_{12}$, which can be obtained from experiment from the transition enthalpy at atmospheric pressure, the melting temperature, the pressure dependence of the transition profile, and the width of the transition.

The change in the free energy of the membrane as a function of temperature and pressure, i.e. the difference in chemical potential between fluid and gel membrane weighted by the fractions of gel and fluid membrane, is given by
\begin{equation}\label{eq:Theory_A8}
\Delta G_{12}=f_{12}\Delta \mu_0^{12}\;.
\end{equation}
Analogously, for the $34$-equilibrium one obtains a free energy $\Delta G_{34}$ with the parameters $\Delta H_0^{34}$, $\Delta S_0^{34}$, $\Delta V_0^{34}$ and $n_{34}$.

\begin{figure}[htbp]
\centering
\includegraphics[width=225pt,height=96pt]{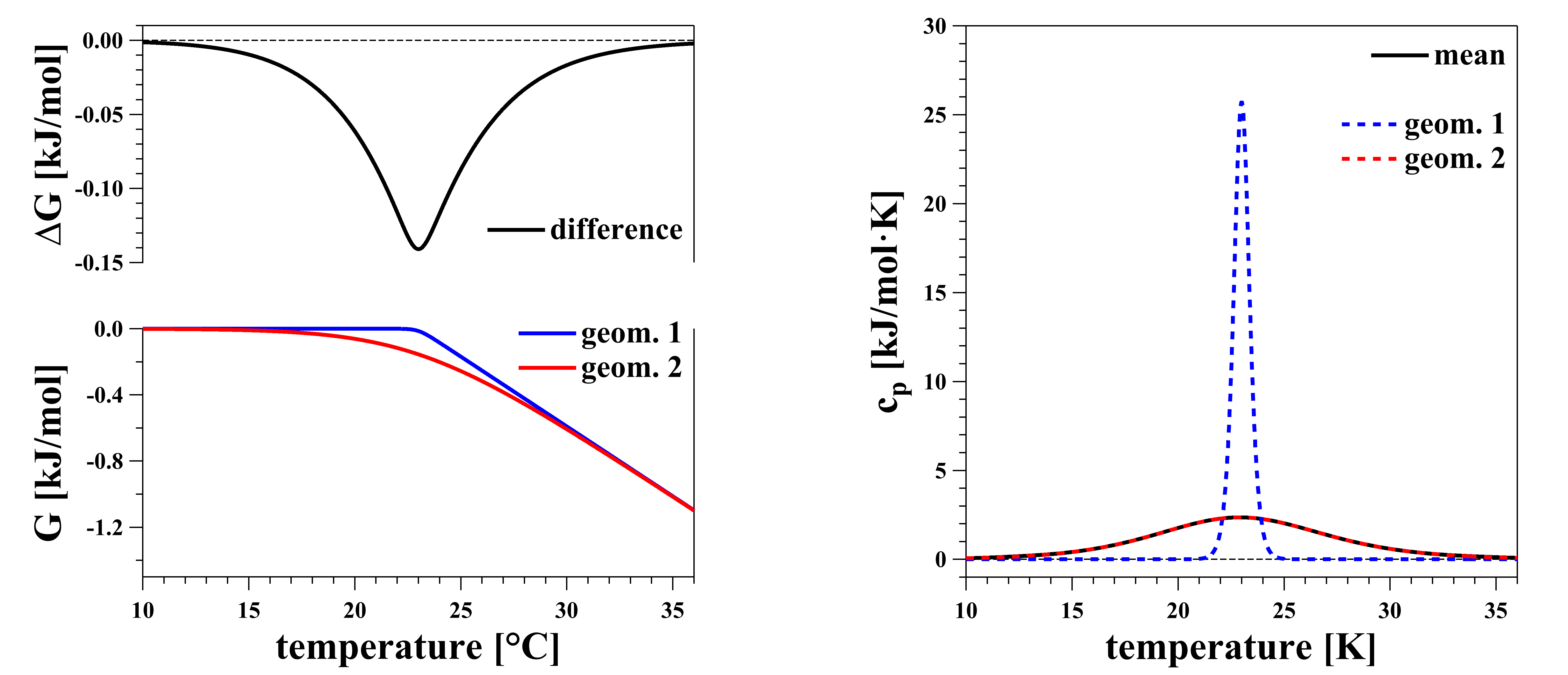}
\caption{Left, bottom: The free energy profiles of two different membrane geometries in the melting regime (blue: geometry 1, $n_{12}=120$. red: geometry 2, $n_{34}=11$). Under these conditions, the free energy of the membrane with lower cooperativity (geometry 2 in red) is always lower than the free energy of the membrane with the higher cooperativity (geometry 1 in blue). Right: Corresponding heat capacity profiles.The blue profile displays the much lower half-width. The red profile is brand and shown as solid line, indicating that this is the profile that one would observe if both geometries are in equilibrium with each other. }
\label{fig_calc_free_energy_cp_nosolvent}
\end{figure}

Let us assume a lipid similar to DMPC with $\Delta H_0^{12}=25 kJ/mol$, $T_m^{12}=296.15$ K ($\Delta S_0^{12}=\Delta H_0^{12}/T_m^{12}$) and a cooperative unit size of $n_{12}=120$ at atmospheric pressure ($p=0$). Then, $\Delta \mu_0^{12}=\Delta H_0^{12}-T\Delta S_0^{12}$. The corresponding free energy profiles is given as the blue profile in Fig. \ref{fig_calc_free_energy_cp_nosolvent} (left, bottom). To demonstrate the role of the cooperativity, let us assume another state of the membrane, with the only difference that the cooperative unit size of $n=11$. This yields the red profile in Fig. \ref{fig_calc_free_energy_cp_nosolvent} (left, bottom). It displays a smaller curvature of the free energy profile. We showed earlier that membrane bending can result in such a broadening of the melting profile \cite{Ivanova2001}. In the following we will assume that the broader melting profiles are associated to higher membrane curvature (as was already suggested in \cite{Schneider1999} and as experiments suggest for the intermediate lipid phase of DMPG \cite{Alakoskela2010}). It can be seen that the red profile is always below the blue profile. Thus, the difference in free energy between red profile and blue profile is always negative (Fig. \ref{fig_calc_free_energy_cp_nosolvent} (left, top)). If two geometries are in equilibrium with each other via a transition in structure, the red profile always corresponds to the more likely state. The heat capacity profiles for the independent melting of the two geometries are shown in Fig. \ref{fig_calc_free_energy_cp_nosolvent} (right). They can be calculated from $\Delta G_{12}$ and $\Delta G_{34}$ by using that $\Delta c_p=-T\cdot (d^2 \Delta G/d T^2)_p$. One finds that a lower cooperative unit size leads to a broader transition profile even if all other parameters are the same.

\subsection{Equilibrium between two geometries}
\label{equilibriumbetweentwogeometries}

\begin{figure}[htbp]
\centering
\includegraphics[width=225pt,height=96pt]{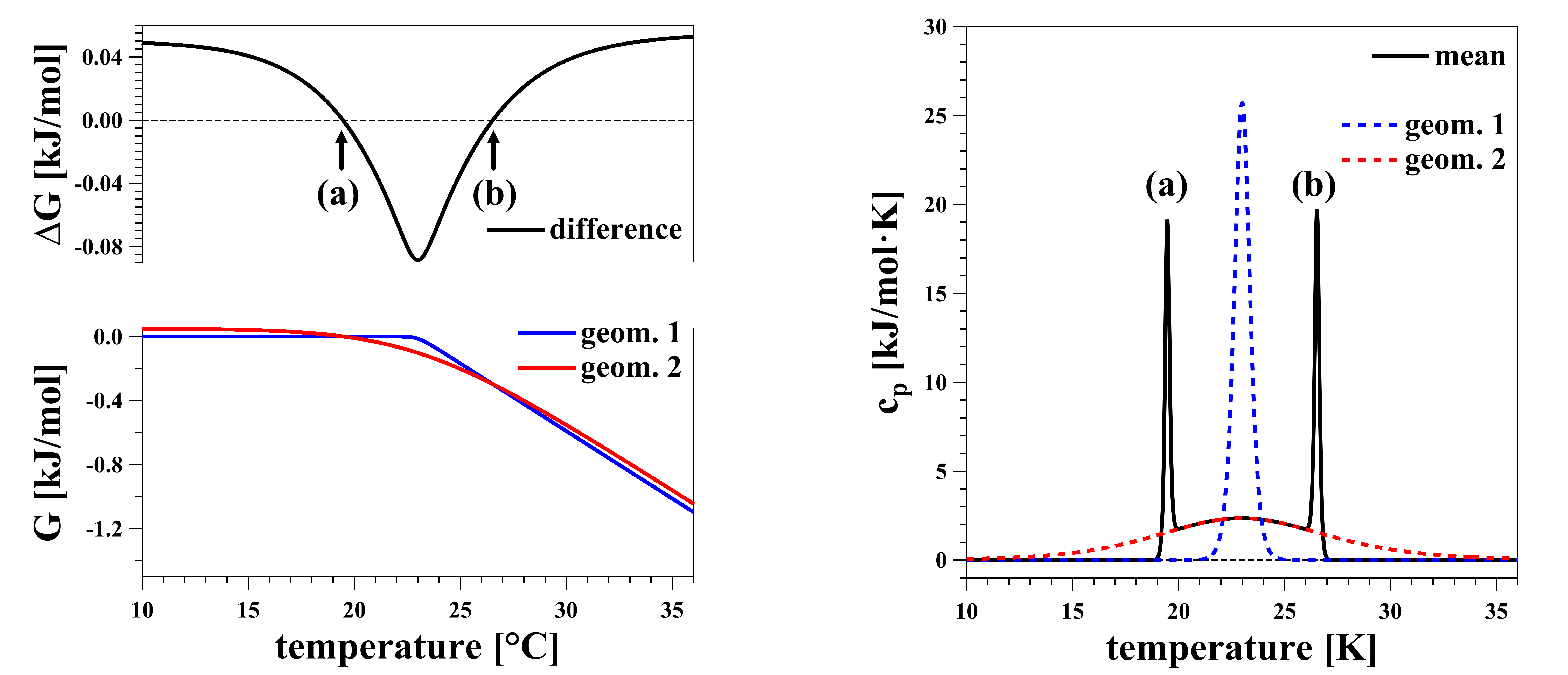}
\caption{Left, bottom: Same free energy profiles as in the the previous figure, but the two geometries interact differently with solvent such that the red profile is shifted by $\Delta \mu_0^{13}=\Delta \mu_0^{24=}50$ J\slash{}mol relative to the blue profile. Further, $\Delta H_0^{12}=\Delta H_0^{34}=25$ kJ\slash{}mol and the melting temperature $T_m^{12}=T_m^{34}=296-15$ K which implies the the melting entropies are the same. The cooperative using sizes are different: $n_{12}=120$, $n_{34}=11$. Now, the two free energy profiles intersect because $\Delta \mu_0^{13}>0$. Left, top: The free energy difference between the two geometries is positive at low and high temperature, while it is negative in the transition range. Right: The corresponding heat capacity profiles. At low and high temperature, the effective heat capacity (in black) follows the blue profile, while in the transition regime it follows the red profile. The geometrical transitions at the intersection points of the free energy lead to sharp heat capacity maxima. They correspond to the peaks labeled (a) and (b) in Fig. \ref{fig01_dmpg_intro}. The analysis was given in \cite{Schneider1999}.}
\label{fig_calc_free_energy_cp_withsolvent_symmetric}
\end{figure}

So far, we have treated the two geometries independent of each other, and we determined their free energies and heat capacity. Let us now consider an equilibrium between geometry 1 and 2, as shown in Fig. \ref{diagram_e}. Let us further assume that the two membrane geometries display a different interaction with the solvent. What we mean here by geometries are two different membranes with $n_{13}$ lipids that display different curvature. In Fig. \ref{fig_calc_free_energy_cp_withsolvent_symmetric} (left) we add a constant free energy $\Delta \mu_0^{13}=\Delta \mu_0^{23}=50$ J\slash{}mol to both gel and fluid state of the membranes of geometry 2, i.e., $\Delta G_{34}\rightarrow \Delta G_{34}+\Delta G_{13}$. The two free energy profiles (in red and blue) now intersect at two temperatures. Below and above the melting regime, geometry 1 (blue line) displays a lower free energy and therefore is the favorable membrane geometry. In the transition regime, the free energy of geometry 2 (red line) is lower than the free energy of geometry 1. Thus, geometry 2 is more stable in the transition regime. At the intersection temperatures one expects transitions between the geometries. The chemical potential difference between the two geometries is given by
\begin{equation}\label{eq:Theory_A10}
\Delta \mu_0^{geom}=(\Delta \mu_0^{34}+\Delta \mu_0^{13})-\Delta \mu_0^{12} \;.
\end{equation} 
Here, the $\mu_{ij}$ are again functions of pressure and temperature.

The equilibrium constant in a transition between the two geometries is given by
\begin{equation}\label{eq:Theory_A11}
K_{geom}=\exp\left(-n_{mem}\frac{\Delta \mu_0^{geom}}{RT}\right) \;.
\end{equation} 
Here, $n_{mem}$ is the overall number of lipids in the membrane under consideration. The fraction of membranes in geometry 2 is given by
\begin{equation}\label{eq:Theory_A11b}
f_{geom}=\frac{K_{geom}}{1+K_{geom}} \;.
\end{equation} 

The measured enthalpy change of a lipid dispersion that may undergo a transition, i.e., the mean enthalpy of the two geometries weighted by their fractions, is given by
\begin{equation}\label{eq:Theory_A12}
\Delta H_{mean}= (1-f_{geom})\Delta H_{12}+f_{geom}\cdot(\Delta H_{13}+\Delta H_{34})\;,
\end{equation} 
where $\Delta H_{13}$ represents a change in solvent interactions when going from geometry 1 to geometry 2.

The heat capacity of the whole system is found after differentiating:
\begin{equation}\label{eq:Theory_A14}
\Delta c_{p, mean}=\left( \frac{d \Delta H_{mean}}{dT}\right)_p\;.
\end{equation} 
For $p=0$, $\Delta \mu_{0}^{13}=50$ J\slash{}mol and otherwise identical parameters as in Fig. \ref{fig_calc_free_energy_cp_nosolvent}, this function is shown in Fig. \ref{fig_calc_free_energy_cp_withsolvent_symmetric} (right, black solid line). It is compared with the melting profiles of the two geometries if they were not in equilibrium with each other (red and blue dashed curves). One can recognize that the $c_p$-profile displays two sharp peaks at the outer limits of the transition, and a broad maximum in between that corresponds to the melting of state 2. The profile is nearly symmetric because we assumed that both geometries display the same melting enthalpy and entropy. While being more detailed and including pressure, the above calculation otherwise resembles that theory presented in \cite{Schneider1999}. This profile is qualitatively similar to the melting of DMPG but quantitatively different in so far as the experimental profiles are not symmetric.

The mean volume change is given by
\begin{equation}\label{eq:Theory_A12b}
\Delta V_{mean}= (1-f_{geom})\Delta V_{12}+f_{geom}\cdot(\Delta V_{13}+\Delta V_{34})\;,
\end{equation} 
where $\Delta V_{13}$ represents a change in volume caused by changing solvent interaction between the two geometries.

By integrating $\Delta c_{p, mean}/T$ we obtain the entropy change of the system
\begin{equation}\label{eq:Theory_A15}
\Delta S_{mean}=\int_{T_0}^{T_1}\frac{\Delta c_{p, mean}}{T}dT\;,
\end{equation} 
and finally the free energy changes
\begin{equation}\label{eq:Theory_A16}
\Delta G_{mean}=\Delta H_{mean} + p\Delta V_{mean} -T\Delta S_{mean}\;,
\end{equation} 
which depend on temperature and pressure.

\begin{figure}[htbp]
\centering
\includegraphics[width=225pt,height=96pt]{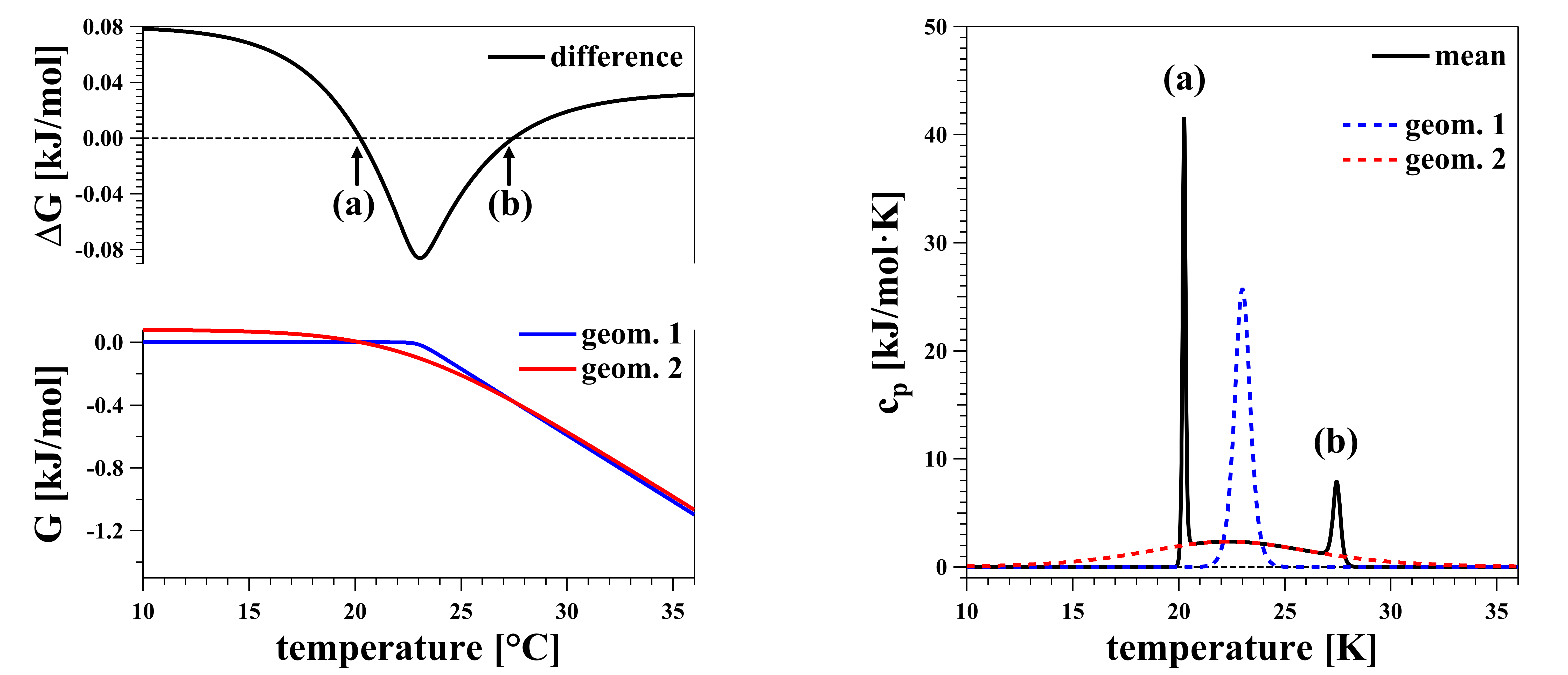}
\caption{Same as the previous two figures but with different solvent interaction in gel and fluid state: $\Delta G_{solv}(gel)=80$ J\slash{}mol$\cdot$K and $G_{solv}(fluid)=25$ J\slash{}mol$\cdot$K. This yields an asymmetric $\Delta c_p$-profile that is quite similar to the experimental profile of DMPG in Fig. \ref{dmpg_pressure_comp_exp_calc} at 1 bar pressure. The sharp peaks of the black profile originate from the structural transitions between the two geometries. They correspond to the peaks labeled (a) and (b) in Fig. \ref{fig01_dmpg_intro}.}
\label{fig_calc_free_energy_cp_withsolvent_asymmetric}
\end{figure}

Experimental melting profiles of DMPG are asymmetric (see Fig. \ref{fig01_dmpg_intro}). In Fig. \ref{fig_calc_free_energy_cp_withsolvent_asymmetric} we show a calculation where we slightly adjusted the parameters. We assume that $\Delta H_0^{34}= \Delta H_0^{12}-55$ J\slash{}mol, $\Delta \mu_0^{13}=80$ J\slash{}mol, and all other parameters are the same as in Fig. \ref{fig_calc_free_energy_cp_withsolvent_symmetric}. A summary of the simulation parameters is given in Table 1. Note that the changes in enthalpy attributed to solvent interactions are very small as compared to the absolute values of the melting enthalpy that mostly originates from chain melting. In the Figure, we see that the free energy difference between the two geometries (left, top) and thus the mean heat capacity profile are no longer symmetric. The heat capacity profile in Fig. \ref{fig_calc_free_energy_cp_withsolvent_asymmetric} in fact looks very similar to that in Fig. \ref{fig01_dmpg_intro}. Table 1 contains 12 parameters of which we set 3 to zero. Three parameters can be deduced from the heat capacity profile at high salt. These are $\Delta H_0^{12}$, $\Delta S_0^{12}$ and $n_{12}$ which correspond to the properties of the blue profile in Figs. \ref{fig_calc_free_energy_cp_withsolvent_symmetric} and \ref{fig_calc_free_energy_cp_withsolvent_asymmetric}). $\Delta H_0^{34}$, $\Delta S_0^{34}$ and $n_{34}$ correspond to the properties of the red profile in Figs. \ref{fig_calc_free_energy_cp_withsolvent_symmetric} and \ref{fig_calc_free_energy_cp_withsolvent_asymmetric}.
The final three parameters control the volume changes during melting and the pressure dependence of the calorimetric profiles. $\Delta H_0^{13}$ determines the distance between the two sharp outer peaks that indicate the temperatures of the structural transitions. $\Delta V_0^{13}$ can be deduced from the pressure dependence of the lower and upper limits of the calorimetric profiles, and $n_{13}$ determines the width of the sharp peaks. This means, all nine parameters reflect experimental properties, i.e. melting enthalpy, entropy and volume of the two geometries (six parameters) and the cooperativities of the two melting transition and the structural transition. This implies that there is no free fitting parameter in the calculation.

\begin{table*}[h!]
\begin{center}
\label{tab:table1}
\begin{tabular}{ |p{2cm}p{1.5cm} || p{2cm}p{1.5cm}|| p{0.5cm}p{0.75cm}|| p{2cm}p{2cm}|}
 \hline
\mbox{enthalpy}   & \mbox{[kJ/mol]}    &  \mbox{entropy} & \mbox{[J/mol $\cdot$K]} & \mbox{coop.} &\mbox{unit}   & \mbox{volume} & \mbox{[m$^3$/mol]}  \\
 \hline
 \hline
$\Delta H_0^{12} $   & 25   & $ \Delta S_0^{12 }$&  84.417  &  n$_{12}$ &  120 & $\Delta V_0^{12 }$&   1.8075$\cdot 10^{-5}$   \\
\hline
$\Delta H_0^{34}-\Delta H_0^{12}$ & -0.055 & $\Delta S_0^{34} - \Delta S_0^{12 }$& 0  & n$_{34}$ &  11  & $\Delta V_0^{34} -\Delta V_0^{12}$ & 0 \\
\hline
$\Delta H_0^{13}$&   0.080  &  $\Delta S_0^{13 }  $ & 0  &n$_{13}$ &  2000 & $\Delta V_0^{13 }$  & - 0.1048$\cdot 10^{-5}$ \\
\hline
\end{tabular}
\caption{Simulation parameters for Figs. \ref{fig_calc_free_energy_cp_withsolvent_asymmetric}, \ref{figure_grabitz_pressure_exp_sim} and \ref{fig_4}. It contains 12 parameters, of which 3 were set to zero. }
\end{center}
\end{table*}


\subsection{Pressure dependence}
\label{pressuredependence}

Calorimetric profiles are pressure-dependent because the chemical potential difference contains a $p\Delta V_0$ term. In the previous section, pressure changes were not considered and therefore the excess volume changes were not needed. However, in eq. (\ref{eq:Theory_A1}) volume changes and pressure are intrinsic parts of the chemical potential differences.

\begin{figure}[htbp]
\centering
\includegraphics[width=225pt,height=100pt]{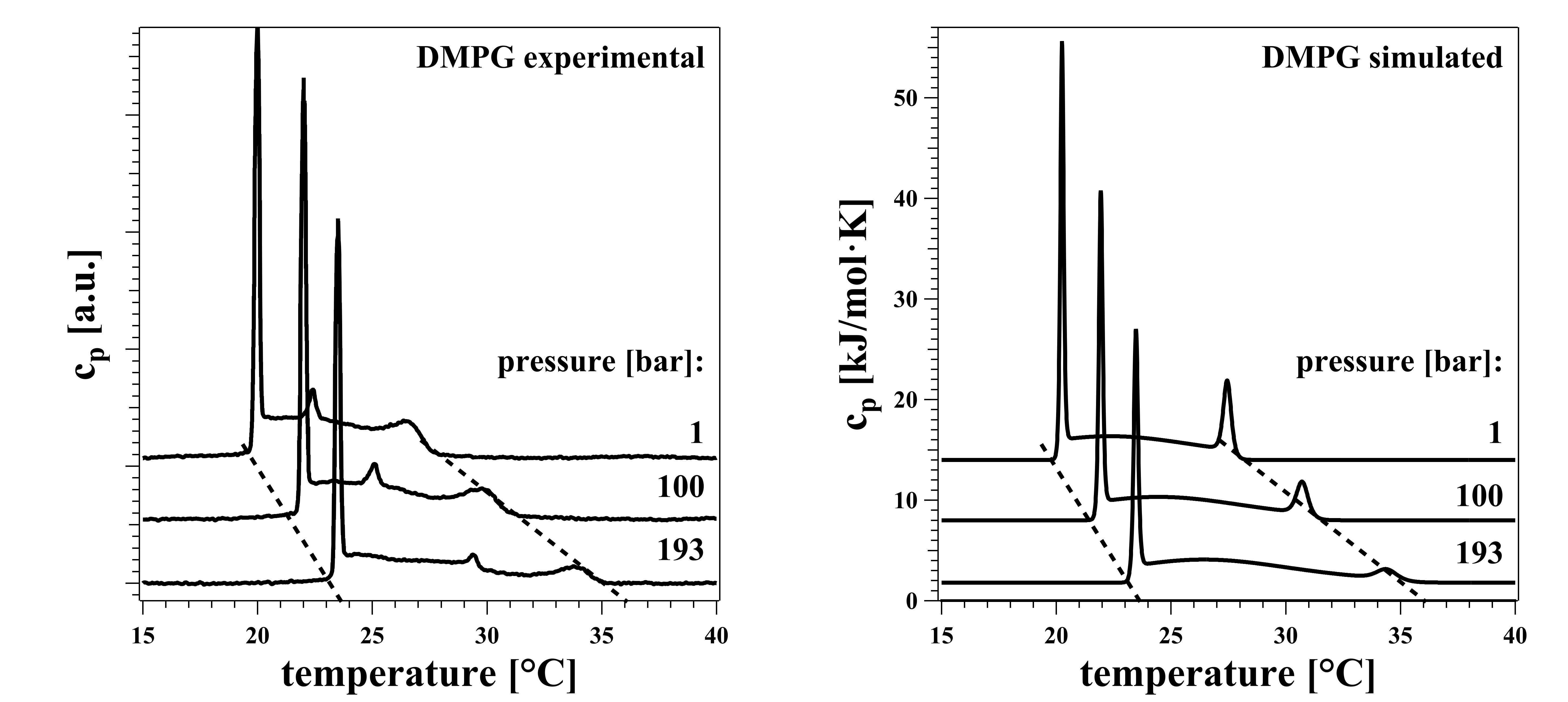}
\caption{Left: Experimental heat capacity profiles of DMPG (10mM NaCl) at three different pressures. Right: Simulation of the heat capacity profiles assuming an equilibrium between two membrane geometries. The dashed lines mark the lower and upper ends of the transitions. They are identical in the left and right hand panel..}
\label{figure_grabitz_pressure_exp_sim}
\end{figure}

Fig. \ref{dmpg_pressure_comp_exp_calc} showed that the pressure dependence of the calorimetric profiles of DMPG cannot be satisfactorily predicted when assuming a proportional relation between volume and enthalpy. However, if we attribute a volume change to the solvent interactions, $\Delta \mu_0^{13}$, then we can actually quite nicely describe the experimental findings. This is shown in Fig. \ref{figure_grabitz_pressure_exp_sim}. The volume changes responsible for the pressure dependence are given in Table 1. Here we chose a volume change of $\Delta V_0^{13}=-0.1048 \cdot 10^{-5}$ m$^3$\slash{}mol (corresponding to about -5.8 \% of the volume change in the melting transitions). The dashed straight lines in the figure mark the lower and upper end of the transition, which are identical in the measured and the calculated curves. This implies that geometry 2 displays a smaller volume associated with the membranes as geometry 1. If one interprets the volume change of associated solvent as the result of an ordering process, one would suspect that geometry 1 (outside of the melting regime) has a more ordered lipid-water interface than geometry 2 in the melting regime and that the change in geometry is partially a consequence of the gain in entropy of interfacial water in the melting regime.

\subsection{Comparison of $\Delta c_p$ with $\alpha$}
\label{comparisonofdeltac_pwithalpha}

\begin{figure}[htbp]
\centering
\includegraphics[width=225pt,height=118pt]{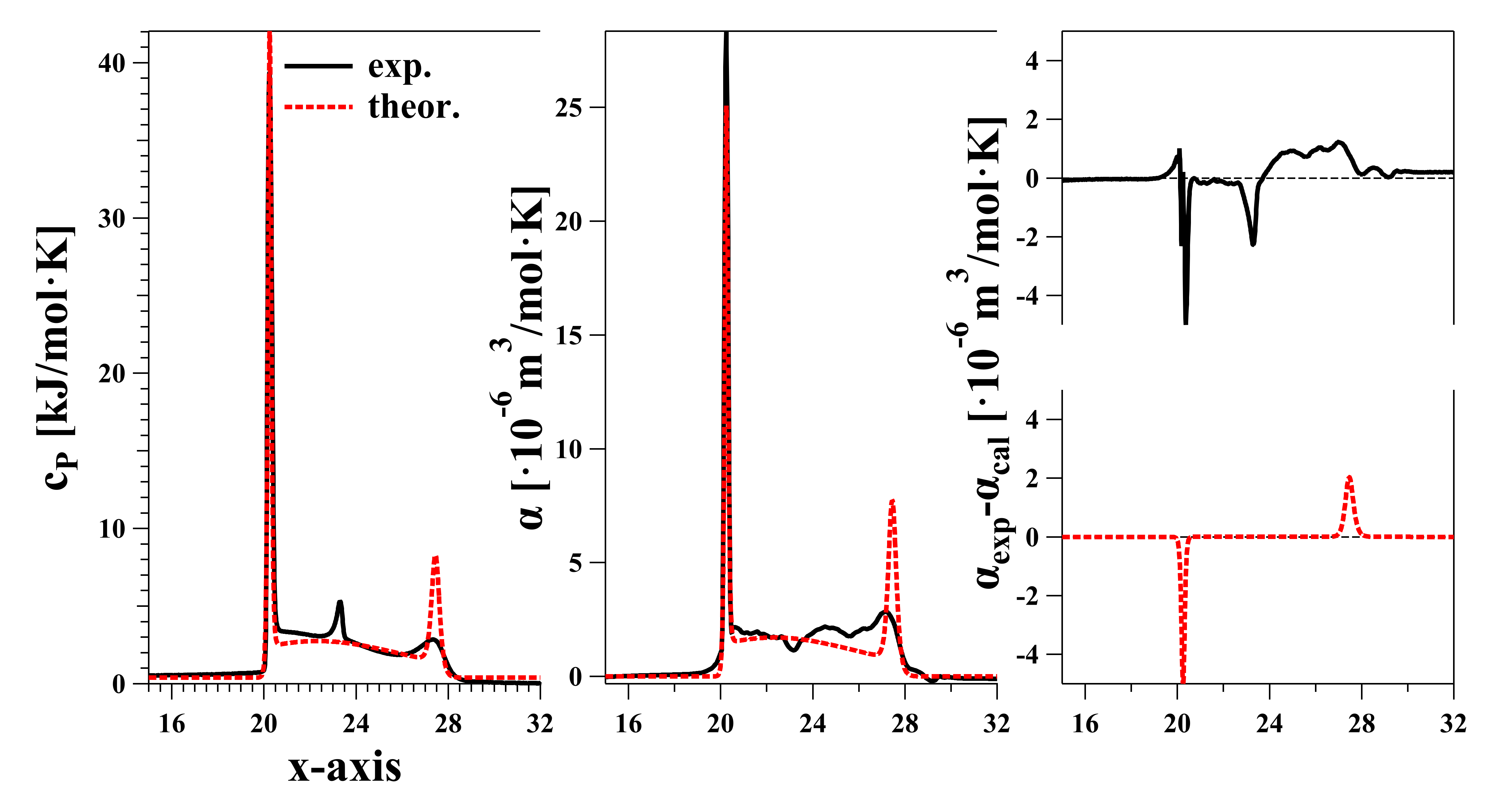}
\caption{Left: Heat capacity $c_p$ of a 5mM DMPG dispersion at 10 mM NaCl and a pH of 7.8 (solid black) compared to a simulation (red dashed, see text for details). Center: Volume expansion coefficient $\alpha$ (measured in a vibrating capillary) compared with the simulation (red dashed, see text). Right: Difference between volume expansion coefficient and heat capacity ($f(T)=\alpha-\gamma_V\cdot c_p$). Both in experiment (black, top) and simulation (red, bottom) this function deviates from a zero baseline showing that the volume expansion coefficient of the lipid dispersion does not display the same temperature dependence than the heat capacity. }
\label{fig_4}
\end{figure}

Fig. \ref{fig_4} shows the differences in the heat capacity (left panel) and the volume expansion coefficient $\alpha$ (center panel) both in experiment and theory. The right hand panel shows the function $f(T)=\alpha-\gamma_V\cdot c_p$. We use $\gamma = 7.23\cdot 10^{-10}$ m$^2$\slash{}N (deduced from the data in Fig. \ref{figure_ebel_dmpg_cp_alpha}). For uncharged lipid membranes, this function would yield a zero baseline. Here, we see that both in experiment and theory, the deviations in the volume change are negative at the onset of the transition, and positive at the upper end of the transition. This confirms the assumption in the derivation of the theory that enthalpy and volume changes during the melting of the two geometries seen separately are proportional, and that the deviations originate from the change in hydration during the transitions in shape. This underlines the importance of solvent interactions for the state of lipid membranes. The experimental and theoretical profiles in the right hand panel deviate somewhat because our theory assumes only two different geometries and does not reproduce the peak in the center. However, both profiles reflect a negative change in $f(T)$ at the lower end of the transition, and a positive change at the upper end of the transition.

\section{Discussion}
\label{discussion}

The peculiar transitions in DMPG were first described by Gershfeld and Nossal \cite{Gershfeld1986, Gershfeld1989}. The increase in viscosity, the appearance of two sharp transition peaks at the lower and upper edge of the melting range and the striking reduction of light diffraction was investigate by \cite{Heimburg1994}. Gershfeld and Nossal focused on the upper sharp peak in the melting profile (that we label here as (b), see Fig. \ref{fig01_dmpg_intro}) and proposed that there exists a critical unilamellar transition that explains this peak. The explanation favored here (which is an extension of a model shown in \cite{Schneider1999}) is very different and involves two different membrane geometries. It seems to be more consistent with the presence a melting regime the extends over several degrees and displays several sharp maxima.

Melting transitions of DMPG extend over several degrees and display three peaks, labelled by (a), (b) and (c) in Fig. \ref{fig01_dmpg_intro}. We call the temperature regime between the two outer peaks the ``intermediate regime'', which is simultaneously the melting regime. It displays a different membrane morphology as indicated by large changes in viscosity, a drastic reduction in light scattering, and obvious changes in electron microscopy. We show here that one can understand the peculiar melting behavior of DMPG quite well when assuming that there exist two geometries of the membrane that both can melt and that are in equilibrium with each other.
Besides experimental enthalpies and entropies, our model also contains cooperative unit sizes. There are two cooperative unit sizes in our model. The first determine the width of the melting peaks of each geometry ($n_{12}$ and $n_{34}$). It corresponds to domain sizes within the membrane plain. The second corresponds to the number of lipids in the macroscopic aggregates ($n_{13}$). This number is higher. Here, we have chosen the cooperative unit sizes such that the experimental $c_p$ profiles are approximately reproduced. The first type of cooperative unit size is a parameter chosen such that for the two geometries melting profiles of different width are obtained. A higher cooperative unit size renders the $c_p$-profiles narrower. Close to the transition, the heat capacity of the more cooperative transition is higher than the of the broad profile and so is its bending elasticity. In this temperature regime, the flat membrane can collapse to the curved structure if that is favored by solvent interactions \cite{Alakoskela2010}.
Thus, one of the geometries exists outside of the melting regime (geometry 1) and the other one in the intermediate melting regime (geometry 2) which melts with lower cooperativity (indicated by the small cooperative unit size of $n_{34}=11$, see table 1). We further assume that a change in the geometry of a membrane is favored by interactions with the solvent. Our analysis represent an extended version of the theory given in \cite{Schneider1999}. In principle, we employ a four-state model with two geometries (1 and 2) and two physical states of each geometry (gel and fluid). We show the major features of the transition profile and its pressure dependence can be reproduced with this approach. When comparing the calorimetric profile in Fig. \ref{fig01_dmpg_intro} with the theoretical profile in Fig. \ref{fig_calc_free_energy_cp_withsolvent_asymmetric} one recognizes that the profiles are quite similar and the major features are the same in both graphs. In particular, there exist two sharp peaks at the lower and upper end of the transition that we labeled (a) and (b) in Figs. \ref{fig01_dmpg_intro} , \ref{fig_calc_free_energy_cp_withsolvent_symmetric} and \ref{fig_calc_free_energy_cp_withsolvent_asymmetric}.
The smaller the salt and lipid concentrations, the broader the melting profile. In our model, this is best reflected by the parameter $Delta mu_13$, which determines the vertical positions of the blue and red profiles in Figs. \ref{fig_calc_free_energy_cp_withsolvent_symmetric} and \ref{fig_calc_free_energy_cp_withsolvent_asymmetric} relative to each other. Shifting the red profile upwards or downwards changes the intersection points of the two free energy profiles, bringing the sharp peaks in the calorimetric profile closer together or pushing them further apart.
There are some fine details in the experimental profiles that are lacking in the calculated profiles. The most prominent difference is that the experimental profile displays a minor sharp peak in the center of the transition (labeled (c) in Fig. \ref{fig01_dmpg_intro}) that is absent in the calculated profile. In our theory, the sharp calorimetric events are linked to cooperative transition in geometry (shape). The presence of a sharp peak in the center suggests that having only two geometries might oversimplify the chain of events. Most probably, a six-state model with three geometries which can each be in gel and fluid state would also reproduce the small central peak. Loew et al. \cite{Loew2011} suggested that there may exist several structures within the melting regime. This is also suggested by Fig. \ref{fig_1b} (left) that shows that the viscosity of the dispersion is different below and above peak (c). Additionally, the viscosity itself displays a minimum at (c). A second difference is that the experimental profiles display a broader peak (b) than calculated, which can also be seen in light scattering data. A further deviation between the experimental and the calculated profile is that in the experimental data the high temperature end peak (b) is broader than in the calculation. This is a consequence of the exact shape of the free energy profile and it might be possible that one can optimize this in the calculations by fine-tuning the parameters.

While we can describe the experimental data quite well (including the pressure dependence), it is not well understood how the intermediate phase in the melting regime of DMPG looks like. The model does not give structural detail besides postulating two distinct geometries of the membrane. It seem obvious from electron microscopy data that DMPG below and above the melting regime forms vesicles\cite{Schneider1999, Kinoshita2008}. However, the intermediate viscous phase with low light scattering is obscure. The fact that the dispersions in the melting regime are practically transparent indicate that no structures exist in the dispersion that have sizes in the range of the wave-length of visible light (several 100 nm). Electron microscopy data show that the intermediate phase does not consist of vesicles but rather of sheet or tube-like structures with pronounced ripple formation in the intermediate regime \cite{Schneider1999, Kinoshita2008}. The intermediate phase does not display easily interpretable scattering profiles in small angle x-ray diffraction (SAXS) and neutron scattering (SANS) profiles \cite{Riske2001, Fernandez2008, Preu2010}. Due to low light scattering and high viscosity and the absence of vesicular structures, \cite{Heimburg1994, Schneider1999} proposed that the intermediate structure is a bicontinous sponge phase. However, it lacks the typical diffraction profile for sponge phases in SAXS and SANS. Further, Alakoskela and Kinnunen \cite{Alakoskela2007} rejected this notion because lipids on the intermediate regime do not macroscopically mix with fluorescence probes indicating that the membranes are not macroscopically connected. Riske, Lamy and others proposed the the intermediate phase displays many defects such as pores and line-like defects \cite{Riske2004, Riske2009, Alakoskela2010, Barroso2012} (depicted in \cite{Riske2004}). Thus, the exact nature of the intermediate phase in DMPG remains a mystery.

For our theory, such structural details are not so relevant. The most important prerequisites for the theory presented here to work are

\begin{enumerate}
\item The membrane geometries outside and inside of the melting regime are different, and they display different curvature. This is supported by the vesicular structures found outside of the melting regime, and rippled sheets found in the intermediate temperature regime \cite{Schneider1999, Kinoshita2008} and the finding of enhanced positive spontaneous curvature by \cite{Alakoskela2010} in this regime.

\item Associated with the change in geometry, the interaction of water with the membranes is different in the melting regime and outside of it. If this were not so, the equilibrium between the two geometries would not depend on lipid and water concentration in such a wide regime.

\end{enumerate}

There is solid experimental support for the change in geometry (point 1). Point 2 is also supported experimentally.

The striking feature of DMPG membranes at low salt concentration is its practically limitless swelling in water \cite{Heimburg1994, Riske2002, Loew2011}. One can see changes in the heat capacity profiles from at least 150mM \cite{Heimburg1994} down to below 100 \textmu M \cite{Riske2002} lipid concentration. This essentially means that water in the lipid dispersion is not free because otherwise addition of more water would not change the transition profiles. Instead, one would see macroscopic phase separation of a lipid and an aqueous phase, which is not present in the melting regime of DMPG at low salt. As shown in the introduction, this corresponds to 300 up to more than 100000 water molecules associated to each lipid. This yields a thickness of the water layer surrounding the membranes of 17 nm - 6 \textmu m, i.e., an ordering effect of water around the membrane that far exceeds the reported interbilayer distance of 2nm in multilamellar DPPC. The long-range ordering effect of DMPG is indeed surprising in its magnitude. It seems to be associated with the charge of DMPG because high salt concentrations abolished the anomalous behavior \cite{Heimburg1994, Riske1997}, as does a change in pH (see Fig. \ref{figure_julia_viscosity}). It is known that the ordering of water at lipid interfaces changes in melting transitions \cite{Schonfeldova2021} but it seems likely that the big effect of water in the intermediate regime is associated to the largely enhanced bending elasticity of membranes in melting transitions \cite{Heimburg1998, Dimova2000}. This assumption is in agreement with the finding from neutron spin-echo spectroscopy that the dynamics of the lipid dispersion is enhanced in the melting regime of DMPG, and the DMPG membranes are softer in the melting regime \cite{Kelley2020}.

\begin{figure}[htbp]
\centering
\includegraphics[width=225pt,height=103pt]{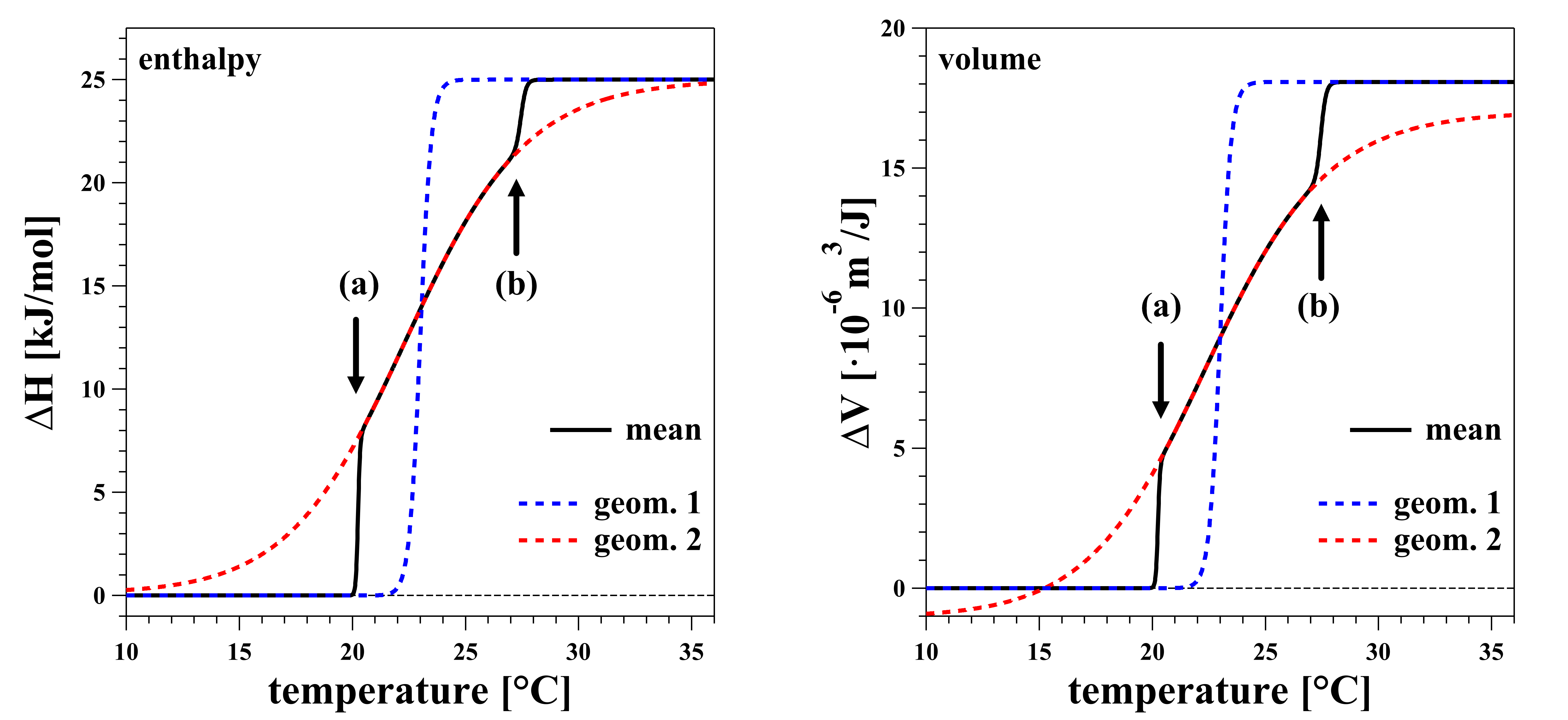}
\caption{Enthalpy (left) and volume changes (right) of the two geometries (red and blue dashed line) and the effective changes (solid black line) as a function of temperature. Due to different volumes of the solvent associated to the membranes, the red profile in the right had panel is shifted downwards. As a consequence, the effective enthalpy and volume profiles are not proportional, while they are proportional when looking at the two geometries separately. }
\label{fig_enthalpy_volume}
\end{figure}

One of the motivations for this article was to investigate whether enthalpy and volume changes are proportional the melting transition of DMPG. For many systems such phospha- tidylcholines, mixtures of two phosphatidylcholines, mixtures of phosphatidylcholines and cholesterol, lung surfactant and even \emph{E.coli} membranes, this was experimentally found \cite{Ebel2001, Muzic2019}. As a consequence, the heat capacity and the compressibility (and other susceptibilities) become proportional functions, too \cite{Heimburg1998, Muzic2019} and it becomes possible to predict the elastic constants of membranes from the heat capacity, which is easy to measure. This convenient fact was used to predict the properties of sound within biomembranes in the soliton theory for nerves.

Here, however, we found deviations of enthalpy and volume changes, which are not exactly proportional functions of temperature. Let us assume that our theory captures the most important features of the transition events in DMPG. Fig.\ref{fig_enthalpy_volume} shows the calculated enthalpy and volume changes of DMPG membranes as a function of temperature as calculated using the parameters from Table \ref{tab:table1}. One can see that the sharp peaks labeled (a) and (b) in Figs. \ref{fig01_dmpg_intro} , \ref{fig_calc_free_energy_cp_withsolvent_symmetric} and \ref{fig_calc_free_energy_cp_withsolvent_asymmetric} are the temperatures where the structural transitions occur and where one finds sudden changes in the enthalpy and the volume. Since the volume changes contain changes in the volume of the associated water, enthalpy and volume changes at these temperature deviate from each other. As a consequence, the proportional relation between enthalpy and volume is not fulfilled at the temperature where structural transitions occur. Therefore, one should be careful when interpreting enthalpy and volume changes in all transitions that involve changes in the interaction with solvent or changes in geometry.

\vspace{0.5cm}
\textbf{Pretransition}

The main point of the present calculation is to understand the coupling of melting transitions with structural transitions. The main reason for the structural change in the melting transition is the change of the elastic constants. Within the transition-range it is much easier to change the membrane curvature because the bending elasticity is higher. If the second geometry has a favorable interaction with the solvent, cooperative changes in membrane geometry can be induced. These two coupled transition will result in melting profiles that displays more than one heat capacity maximum.

\begin{figure}[htbp]
\centering
\includegraphics[width=225pt,height=209pt]{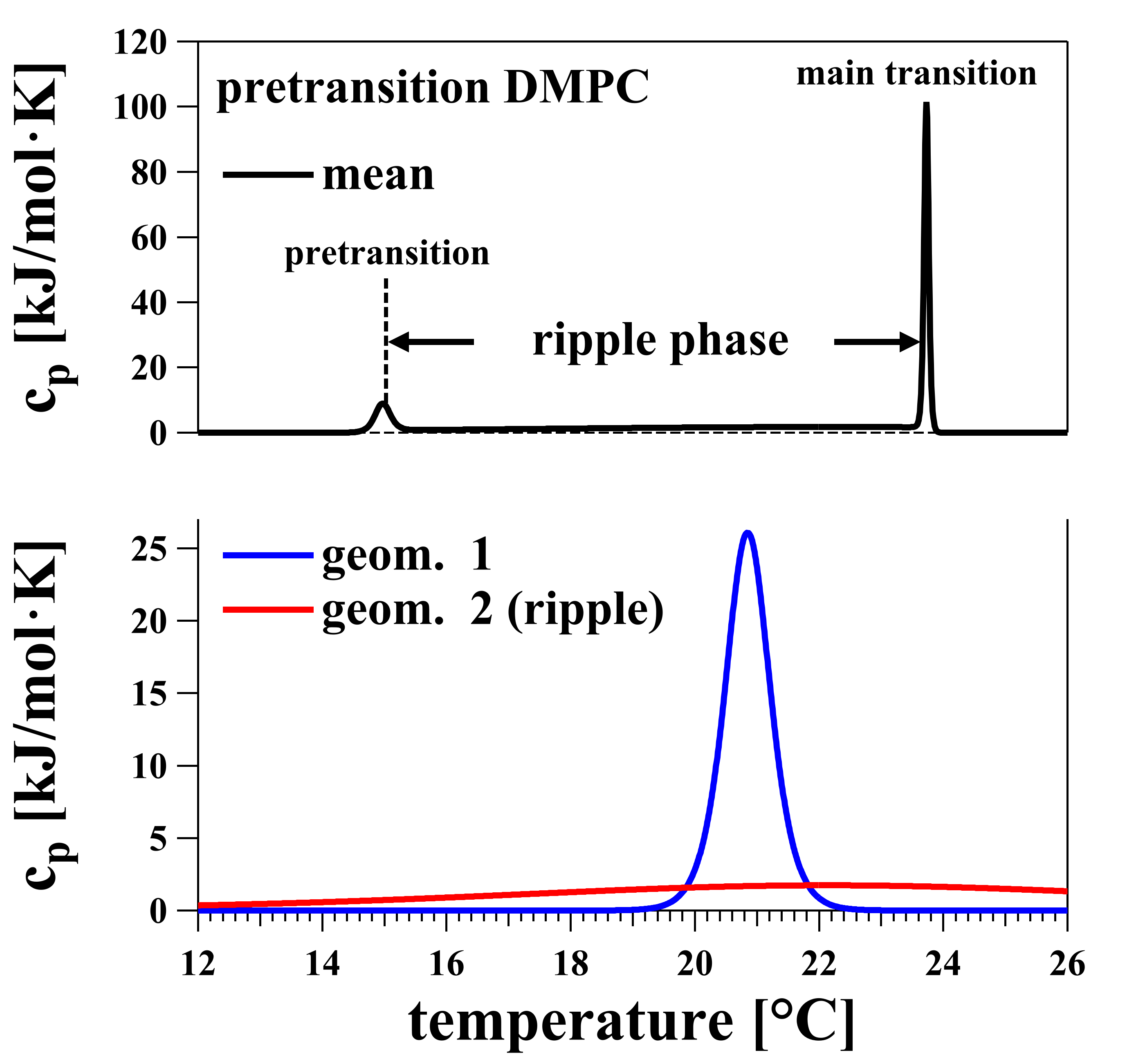}
\caption{With the present theory, one can also simulate $c_p$-profiles that strikingly resemble the pre- and main-transition of DMPC membranes. Simulation parameters are given in Table 2. See text for details.}
\label{fig_dmpc_ripple}
\end{figure}

\begin{table*}[h!]
\begin{center}
\label{tab:table2}
\begin{tabular}{ |p{2cm}p{1.5cm} || p{2cm}p{1.5cm}|| p{0.5cm}p{0.75cm}|}
 \hline
\mbox{enthalpy}   & \mbox{[kJ/mol]}    &  \mbox{entropy} & \mbox{[J/mol $\cdot$K]} & \mbox{coop.} &\mbox{unit} \\
 \hline
 \hline
$\Delta H_0^{12}$   & 25    & $ \Delta S_0^{12 }$&  85.03 &  n$_{12}$ &  120\\
\hline
$\Delta H_0^{34}-\Delta H_0^{12}$ & 0.120 & $\Delta S_0^{34} - \Delta S_0^{12 }$& 0  & n$_{34}$ &  8\\
\hline
$\Delta H_0^{13}$&   0.035  &  $\Delta S_0^{13 }  $ & 0  & n$_{13}$ &  3000 \\
 \hline
\end{tabular}
\caption{Simulation parameters used for generating the pretransition of DMPC shown in Fig. \ref{fig_dmpc_ripple}.  Volume changes are not indicated.}
\end{center}
\end{table*}

Nothing in the present analysis is special for DMPG and one might be tempted to search for other examples for a coupling between melting and geometry changes. One of the obvious candidates for such behavior is the ripple phase the can be observed below the main transition of most phosphatidylcholine (e.g., DMPC and DPPC). For DMPC, one finds a low enthalpy pretransition at around 15$^\circ$C and a main transition at 24$^\circ$C. The total enthalpy change is about 25 kJ\slash{}mol, of which around 4 kJ\slash{}mol originate from the pretransition. As in the DMPG membranes described above, the pretransition displays a different pressure dependence than the main transition \cite{Ebel2001}. In between these two transition peaks, the membranes display periodic undulations of the membrane surface. This phase is called the `ripple' phase. In most of the literature, these are just considered two different independent transitions. However, Heimburg \cite{Heimburg2000} suggested that these two transitions are just part of a continuous melting process and that the ripple phase already contains partially melted regions. The heat capacity does not return to the baseline in between the two peaks. This was confirmed in molecular dynamics simulations of the ripple phase, e.g., by \cite{DeVries2005} and by experimental work from \cite{Riske2009}. Below, we show that the heat capacity profile of DMPC melting can be approximated with the theory presented here by adjusting the solvent interactions (see Table 2). Fig. \ref{fig_dmpc_ripple} shows the results of the calculation, which reproduces relative height and distance of the two peaks, and which displays a $\Delta c_p$ that does not return to zero in between the two peaks. Below the pretransition and above the main transition, the membrane is in the flat state (blue line) without ripples. In between pre- and main transition, the membrane is in its ripple phase (red line). The heat capacity of the ripple state in Fig. \ref{fig_dmpc_ripple} is not zero above the main transition. However, at these temperatures, the membrane is in its flat state (blue) and the heat capacity returns to the base line. The chain of events is analogous to those shown in Fig. \ref{fig_calc_free_energy_cp_withsolvent_asymmetric}.

Since we do not present the pressure dependence of the pretransition here, we did not assign the volume changes. However, this could be done by using the values in \cite{Ebel2001}. The main difference between the values in Tables 1 and 2 is that $\Delta H_0^{34}-\Delta H_0^{12}$ is positive while it was negative in DMPG. $\Delta H_0^{13}$ is smaller than in DMPG, which indicates that the solvent interactions in DMPG are larger than in DMPC. These changes cause that the first peak (the pretransition) is smaller than the second peak (the main transition). We see in this calculation that the sharpness of the major peaks may not arise from a high cooperativity of two separate melting events but rather from the high cooperativity of the structural changes, which are disregarded in most Monte-Carlo simulations (e.g., \cite{Mouritsen1992a, Sugar1994, Heimburg1996a, Ivanova2001}) or molecular dynamics simulations (e.g., \cite{Thomas2022}).
Other transitions that one might treat with this approach include the sub-main transition discussed by K. J\o rgensen \cite{Jorgensen1995}. He showed that in several phosphatidylcholine one finds a minor peak between the pretransition and the main transition. This resembles the three peaks (a), (b) and (c) of DMPG shown in Fig. \ref{fig01_dmpg_intro}. A further transition that displays more than one peak is the split peak in DPPC large unilamellar vesicles \cite{Heimburg1998, Ebel2001}.

One might wonder if the phenomena described here could be relevant in biology. Cells contain approximately 70\% water \cite{Ling2004}. Let us assume that the remaining matter in a cell consists of a small protein, such as cytochrome c, which has a diameter of around 3 nm. This would correspond to around 1,600 water molecules per protein. Assuming a diameter of about 0.3 nm for a water molecule, this corresponds to an aqueous layer of about 1 nm around each protein. This means that the water layer is significantly smaller than the protein's diameter, such that, under biological conditions, molecules nearly touch each other. Since there are also other molecules and membrane surfaces in cells, the exact thickness of the aqueous interfacial layer may vary. If one assumes that all matter in a cell consists of lipid membranes, the aqueous layer had about 2-3 nm on each side of the bilayer. Thus, the above example provides an indication of the order of magnitude of 1--3 nm aqueous layer on each surface. Thus, cells exist under low water conditions where molecules are crowded \cite{Rivas2016}. The biological membrane is close to a phase transition. It undergoes structural changes during endocytosis and exocytosis, for example in cellular transport and secretion or signal transmission in synapses, but also during the action potential \cite{Heimburg2005c}. These processes may depend heavily on the competitions of different surfaces and membrane geometries for water.

The message of the present work is that the interactions of membrane with solvent close to transitions can facilitate such fission and fusion processes in living cells and nerves.

Summarizing, we conclude that the coupling of geometry with melting transitions might be a generic phenomenon that is commonly found. It is not special for DMPG even though it is to our knowledge the system where such changes are most obvious. This suggests that the interactions of membranes with the surrounding solvent might not be treated with the attention that they deserve. It may emerge that membranes and the solvent are not two separate phases, but that the membrane-solvent complex is a macroscopic phase in itself. This is clearly the case for DMPG.






\begin{thebibliography}{10}
\expandafter\ifx\csname url\endcsname\relax
  \def\url#1{\texttt{#1}}\fi
\expandafter\ifx\csname urlprefix\endcsname\relax\def\urlprefix{URL }\fi
\expandafter\ifx\csname href\endcsname\relax
  \def\href#1#2{#2} \def\path#1{#1}\fi

\bibitem{Heerklotz2002b}
H.~Heerklotz, J.~Seelig, Application of pressure perturbation calorimetry to
  lipid bilayers., Biophys.\ J. 82 (2002) 1445--1452.

\bibitem{Loew2011}
C.~Loew, K.~A. Riske, M.~T. Lamy, J.~Seelig., Thermal phase behavior of {DMPG}
  bilayers in aqueous dispersions as revealed by $^2${H}- and $^31${P-NMR},
  Langmuir 27 (2011) 10041--10049.

\bibitem{Muzic2019}
T.~Mu\v{z}i\'{c}, F.~Tounsi, S.~B. Madsen, D.~Pollakowski, M.~Konrad,
  T.~Heimburg, Melting transitions in biomembranes, Biochim.\ Biophys.\ Acta
  1861 (2019) 183026.

\bibitem{Heimburg2005c}
T.~Heimburg, A.~D. Jackson, On soliton propagation in biomembranes and nerves,
  Proc.\ Natl.\ Acad.\ Sci.\ USA 102 (2005) 9790--9795.

\bibitem{Heimburg2007a}
T.~Heimburg, Thermal biophysics of membranes, Wiley VCH, Berlin, Germany, 2007.

\bibitem{Heimburg1998}
T.~Heimburg, Mechanical aspects of membrane thermodynamics. {E}stimation of the
  mechanical properties of lipid membranes close to the chain melting
  transition from calorimetry, Biochim.\ Biophys.\ Acta 1415 (1998) 147--162.

\bibitem{Heimburg2012}
T.~Heimburg, The capacitance and electromechanical coupling of lipid membranes
  close to transitions. the effect of electrostriction., Biophys.\ J. 103
  (2012) 918--929.

\bibitem{Heimburg2000}
T.~Heimburg, A model for the lipid pretransition: Coupling of ripple formation
  with the chain-melting transition, Biophys.\ J. 78 (2000) 1154--1165.

\bibitem{Heimburg1994}
T.~Heimburg, R.~L. Biltonen, Thermotropic behavior of
  dimyristoylphosphatidylglycerol and its interaction with cytochrome c,
  Biochemistry 33 (1994) 9477--9488.

\bibitem{Gershfeld1986}
N.~L. Gershfeld, W.~F. Stevens, R.~J. Nossal., Equilibrium studies of
  phospholipid bilayer assembly. {C}oexistence of surface bilayers and
  unilamellar vesicles., Faraday Discuss.\ Chem.\ Soc. 81 (1986) 19--28.

\bibitem{Gershfeld1989}
N.~L. Gershfeld, The critical unilamellar lipid state: a perspective for
  membrane bilayer assembly, Biochim.\ Biophys.\ Acta 988 (1989) 335--350.

\bibitem{Riske1997}
K.~A. Riske, M.~J. Politi, W.~F. Reed, M.~T. Lamy-Freund, Temperature and ionic
  strength dependent light scattering of dmpg dispersions, Chem.\ Phys.\ Lett.
  89 (1997) 31--44.

\bibitem{Schneider1999}
M.~F. Schneider, D.~Marsh, W.~Jahn, B.~Kloesgen, T.~Heimburg, Network formation
  of lipid membranes: Triggering structural transitions by chain melting,
  Proc.\ Natl.\ Acad.\ Sci.\ USA 96 (1999) 14312--14317.

\bibitem{Riske1999}
K.~A. Riske, O.~R. Nascimento, M.~Peric, B.~L. Bales, M.~T. Lamy-Freund,
  Probing dmpg vesicle surface with a cationic aqueous soluble spin label,
  Biochim.\ Biophys.\ Acta 1418 (1997) 133--146.

\bibitem{LamyFreund2003}
M.~T. Lamy-Freund, K.~A. Riske, The peculiar thermo-structural behavior of the
  anionic lipid {DMPG}, Chem.\ Phys.\ Lipids 122 (2003) 19--32.

\bibitem{Riske2003}
K.~A. Riske, R.~M. Fernandez, O.~R. Nascimento, B.~L. Bales, M.~T.
  Lamy-Freund., {DMPG} gel--fluid thermal transition monitored by a
  phospholipid spin labeled at the acyl chain end., Chem.\ Phys.\ Lipids 124
  (2003) 69--80.

\bibitem{Riske2003b}
K.~A. Riske, H.~D\"obereiner, M.~T. Lamy-Freund., Comment on ``{G}el-{F}luid
  transition in dilute versus concentrated {DMPG} aqueous dispersions.'', J.\
  Phys.\ Chem.\ B 107 (2003) 5391--5392.

\bibitem{Kinoshita2008}
M.~Kinoshita, S.~Kato, H.~Takahashi, Effect of bilayer morphology on the subgel
  phase formation., Chem.\ Phys.\ Lipids 151 (2008) 30--40.

\bibitem{Fernandez2008}
R.~M. Fernandez, K.~A. Riske, L.~Amaral, R.~Itri, M.~T. Lamy., Influence of
  salt on the structure of {DMPG} studied by {SAXS} and optical microscopy.,
  Biochim.\ Biophys.\ Acta 1778 (2008) 907--916.

\bibitem{Shen2008}
Y.~Shen, J.~Hao, H.~Hoffmann, Z.~Wu., Reversible phase transition from vesicles
  to lamellar network structures triggered by chain melting, Soft Matter 4
  (2008) 805--810.

\bibitem{Riske2009}
K.~A. Riske, R.~P. Barroso, C.~C. Vequi-Suplicy, R.~Germano, V.~B. Henriques,
  M.~Lamy., Lipid bilayer pre-transition as the beginning of the melting
  process., Biochim.\ Biophys.\ Acta 1788~(2009. Biochim. Biophys. Acta (BBA) -
  Biomembr. 1788:954--963.) (2009) 954--963.

\bibitem{Riske2009b}
K.~A. Riske, L.~Q. Amaral, M.~T. Lamy., Extensive bilayer perforation coupled
  with the phase transition region of an anionic phospholipid, Langmuir 25
  (2009) 10083--10091.

\bibitem{Spinozzi2010}
F.~Spinozzi, L.~Paccamiccio, P.~Mariani, , L.~Q. Amaral., Melting regime of the
  anionic phospholipid {DMPG}: {N}ew lamellar phase and porous bilayer model.,
  Langmuir 26 (2010) 6484--6493.

\bibitem{Alakoskela2010}
J.-M. Alakoskela, M.~J. Parry, P.~K.~J. Kinnunen., The intermediate state of
  {DMPG} is stabilized by enhanced positive spontaneous curvature., Langmuir 26
  (2010) 4892--4900.

\bibitem{Barroso2010}
R.~P. Barroso, K.~A. Riske, V.~B. Henriques, M.~T. Lamy., Ionization and
  structural changes of the {DMPG} vesicle along its anomalous gel-fluid
  phase transition: {A} study with different lipid concentrations., Langmuir 26
  (2010) 13805--13814.

\bibitem{Henriques2011}
V.~B. Henriques, R.~Germano, M.~T. Lamy, M.~N. Tamashiro., Phase transitions
  and spatially ordered counterion association in ionic-lipid membranes:
  {T}heory versus experiment, Langmuir 27 (2011) 13130--13143.

\bibitem{Ito2015}
A.~S. Ito, A.~P. Rodrigues, W.~M. Pazin, M.~Barioni., Fluorescence
  depolarization analysis of thermal phase transition in {DPPC} and {DMPG}
  aqueous dispersions, J.\ Lumin. 158 (2015) 153--159.

\bibitem{Kelley2020}
E.~G. Kelley, M.~Nagao, P.~D. Butler, L.~Porcar, B.~Farago., Enhanced dynamics
  in the anomalous melting regime of {DMPG} lipid membranes., Struct.\ Dyn. 7
  (2020) 054704.

\bibitem{Schonfeldova2021}
T.~Sch\"onfeldov\'a, P.~Piller, F.~Kovacik, G.~Pabst, H.~I. Okur, , S.~Roke.,
  Lipid melting transitions involve structural redistribution of interfacial
  water., J.\ Phys.\ Chem.\ B 125 (2021) 12457--12465.

\bibitem{Gershfeld2023}
N.~L. Gershfeld, R.~Nossal., Critical point for membrane bilayer formation.,
  Biochim.\ Biophys.\ Acta 1865 (2023) 184116.

\bibitem{Kodama1998}
M.~Kodama, J.~Nakamura, T.~Miyata, H.~Aoki, The behaviour of water molecules
  associated with structural changes in negatively charged
  phosphatidyl-glycerol assemblies as studied by dsc., J.\ Therm.\ Anal.\
  Calorim. 51 (1998) 91--104.

\bibitem{Riske2002}
K.~A. Riske, H.-G. D\"obereiner, M.~T. Lamy-Freund, Gel-fluid transition in
  dilute versus concentrated dmpg aqueous dispersions, J.\ Phys.\ Chem.\ B 106
  (2002) 239--246.

\bibitem{Ebel2001}
H.~Ebel, P.~Grabitz, T.~Heimburg, Enthalpy and volume changes in lipid
  membranes. i. the proportionality of heat and volume changes in the lipid
  melting transition and its implication for the elastic constants, J.\ Phys.\
  Chem.\ B 105 (2001) 7353--7360.

\bibitem{Peters2017}
J.~Peters, J.~Marion, F.~J. Becher, M.~Trapp, T.~Gutberlet, D.~J. Bicout,
  T.~Heimburg, Thermodynamics of lipid multi-lamellar vesicles in presence of
  sterols at high hydrostatic pressure., Sci.\ Rep. 7 (2017) 15339.

\bibitem{Grabitz2002}
P.~Grabitz, V.~P. Ivanova, T.~Heimburg, Relaxation kinetics of lipid membranes
  and its relation to the heat capacity, Biophys.\ J. 82 (2002) 299--309.

\bibitem{Trauble1976}
H.~Tr\"{a}uble, M.~Teubner, P.~Woolley, H.~Eibl, Electrostatic interactions at
  charged lipid membranes. i. {E}ffects of p{H} and univalent cations on
  membrane structure., Biophys.\ Chem. 4 (1976) 319--342.

\bibitem{Ebel1999}
H.~Ebel, Kalorische und strukturelle {U}mwandlungen von {L}ipid-{M}embranen:
  {K}opplung von w\"armekaapazit\"at und volumenexpansionskoeffizienten,
  Master's thesis, {U}niversity of {G}\"ottingen, masters Thesis (1999).

\bibitem{Preu2010}
J.~Preu, Structural changes in charged membrane model systems, Ph.D. thesis,
  {U}niversity of {C}openhagen., Copenhagen (April 2010).

\bibitem{Grabitz2001}
P.~Grabitz, Dynamische {V}org\"ange in {P}hospholipidmembranen.
  relaxationsprozesse und viskosit\"at in der n\"ahe der schmelzumwandlung.,
  Master's thesis, {U}niversity of {G}\"ottingen.

\bibitem{Heimburg1999}
T.~Heimburg, B.~Angerstein, D.~Marsh, Binding of peripheral proteins to mixed
  lipid membranes: Effect of lipid demixing upon binding., Biophys.\ J. 76
  (1999) 2575--2586.

\bibitem{Wang2018}
T.~Wang, T.~Mu\v{z}i\'{c}, A.~D. Jackson, T.~Heimburg, The free energy of
  biomembrane and nerve excitation and the role of anesthetics, Biochim.\
  Biophys.\ Acta 1860 (2018) 2145--2153.

\bibitem{Ivanova2001}
V.~P. Ivanova, T.~Heimburg, A histogram method to obtain heat capacities in
  lipid monolayers, curved bilayers and membranes containing peptides, Phys.\
  Rev.\ E 63 (2001) 041914.
  
\bibitem{Alakoskela2007}
J.-M.~I. Alakoskela and P.~K.~J.~Kinnunen, Thermal phase behavior of DMPG: The 
exclusion of continuous network and dense aggregates., Langmuir 23 (2007) 4203--4213 

\bibitem{Riske2001}
K.~A. Riske, L.~Q. Amaral, M.~T. Lamy-Freund., Thermal transitions of {DMPG}
  bilayers in aqueous solution: {SAXS} structural studies., Biochim.\ Biophys.\
  Acta 1511 (2001) 297--308.

\bibitem{Riske2004}
K.~A.~Riske, L.~Amaral, H.-G. D\"obereiner, M.~Lamy., Mesoscopic structure
  in the chain-melting regime of anionic phospholipid vesicles, Biophys.\ J. 86
  (2004) 3722--3733.

\bibitem{Barroso2012}
R.~P. Barroso, K.~R. Perez, I.~M. Cuccovia, M.~T. Lamy., Aqueous dispersions of
  {DMPG} in low salt contain leaky vesicles., Chem.\ Phys.\ Lipids 165 (2012)
  169--177.

\bibitem{Dimova2000}
R.~Dimova, B.~Pouligny, C.~Dietrich, Pretransitional effects in
  dimyristoylphosphatidylcholine vesicle membranes: Optical dynamometry study,
  Biophys.\ J. 79 (2000) 340--356.

\bibitem{DeVries2005}
A.~H. de~Vries, S.~Yefimov, A.~E. Mark, S.~J. Marrink, Molecular structure of
  the lecithin ripple phase, Proc.\ Natl.\ Acad.\ Sci.\ USA 102 (2005)
  5392--5396.

\bibitem{Mouritsen1992a}
O.~G. Mouritsen, J.~H.Ipsen, K.~J{\o}rgensen, M.~M. Sperotto, Z.~Zhang,
  E.~Corvera, D.~P. Fraser, M.~J. Zuckermann, Computer simulation of phase
  transitions in natures preferred liquid crystal: the lipid bilayer membrane,
  in: G.~F. Luckhurst (Ed.), Computer simulation of liquid crystals, Kluwer
  Academic Publishers, The Netherlands, 1992.

\bibitem{Sugar1994}
I.~P. Sugar, R.~L. Biltonen, N.~Mitchard, Monte {C}arlo simulations of
  membranes: phase transition of small unilamellar
  dipalmitoylphosphatidylcholine vesicles, Methods Enzymol. 240 (1994)
  569--593.

\bibitem{Heimburg1996a}
T.~Heimburg, R.~L. Biltonen, A {M}onte {C}arlo simulation study of
  protein-induced heat capacity changes, Biophys.\ J. 70 (1996) 84--96.

\bibitem{Thomas2022}
N.~Thomas, A.~Agrawal, A lateral electric field inhibits gel-to-fluid
  transition in lipid bilayers, Soft Matter 18 (2022) 6437--6442.

\bibitem{Jorgensen1995}
K.~J{\o}rgensen, Calorimetric detection of a sub-main transition in long-chain phosphatidylcholine lipid bilayers, Biochim.\ Biophys.\ Acta 1240 (1995)
  111--114.

\bibitem{Ling2004}
G.~Ling, What determines the normal water content of a living cell?, Physiol.\
  Chem.\ Phys.\ Med.\ NMR. 36 (2004) 1--19.

\bibitem{Rivas2016}
G.~Rivas, A.~P. Minton, Macromolecular crowding in vitro, in vivo, and in
  between., Trends Biochem.\ Sci. 41 (2016) 970--981.

\end{thebibliography}
\end{document}